\documentclass[jgrga]{AGUTeX}
\usepackage{amsmath}
\usepackage{amsfonts}
\usepackage{amssymb}
\usepackage{graphicx}
\usepackage{verbatim}

\authorrunninghead{DIMANT AND OPPENHEIM}

\titlerunninghead{M-I COUPLING VIA TURBULENT CONDUCTANCE}

\authoraddr{Y. S. Dimant,
Center for Space Physics, Boston University, 725 Commonwealth Ave., Boston, MA 02215, USA.
(dimant@bu.edu)}

\begin{document}

\title{Magnetosphere-Ionosphere Coupling Through \textit{E}-region Turbulence: Anomalous Conductivities and Frictional Heating}

\author{Y. S. Dimant and M. M. Oppenheim}

\begin{abstract}

Global magnetospheric MHD codes using ionospheric conductances
based on laminar models systematically overestimate the
cross-polar cap potential during storm time by up to a factor
of two. At these times, strong DC electric fields penetrate to
the \textit{E} region and drive plasma instabilities that
create turbulence. This plasma density turbulence induces
non-linear currents, while associated electrostatic field
fluctuations result in strong anomalous electron heating. These
two effects will increase the global ionospheric conductance.
Based on the theory of non-linear currents developed in the
companion paper, this paper derives the correction factors
describing turbulent conductivities and calculates turbulent
frictional heating rates. Estimates show that during strong
geomagnetic storms the inclusion of anomalous conductivity can
double the total Pedersen conductance. This may help explain
the overestimation of the cross-polar cap potentials by
existing MHD codes. The turbulent conductivities and frictional
heating presented in this paper should be included in global
magnetospheric codes developed for predictive modeling of space
weather.

\end{abstract}

\begin{article}

\section{Introduction}

In the companion paper \citep{Dimant:Magnetosphere2011_Budget},
we have developed a theoretical description of
magnetosphere-ionosphere (MI) coupling through electrostatic
plasma turbulence in the lower ionosphere in the region where
field-aligned magnetospheric currents close and dissipate
energy. In global MHD computer codes intended for predictive
modeling of space weather, the entire ionosphere plays the role
of the inner boundary condition. The ionospheric conductances
employed in these codes are usually based on simple laminar
models of ionospheric plasma modified by the effects of
precipitating particles. These global magnetospheric MHD codes
with normal conductances often overestimate the cross-polar cap
potential during magnetic (sub)storms up to a factor of two
\citep[e.g.,][]{winglee:1997,raeder:1998,raeder.monograph:2001,
Siscoe:Hill2002,Ober:Testing2003,
Merkin:Global2005,Merkin:Anomalous2005,guild:2008a,
Merkin:Predicting2007,wang.h:2008}.
At the same time, strong DC electric field penetrates to the
\textit{E} region, roughly between 90 and 130~km of altitude,
and drives plasma instabilities. The \textit{E}-region
instabilities create turbulence that consists of density
perturbations coupled to electric field modulations. This
turbulence causes anomalous conductivity which could account
for the discrepancy.

Anomalous conductivity manifests itself in two ways. One is
associated with average anomalous heating by turbulent electric
fields, and the other is due to a net non-linear current (NC)
formed by plasma density turbulence. While the former effect
has been considered before in detail (see references below),
the latter has only been discussed but never quantitatively
investigated with application to ionospheric conductivity. This
is the main objective of this paper. In the two following
paragraphs, we outline each of these effects.

Small-scale fluctuations generated by the \textit{E}-region
instabilities can cause enormous anomalous electron heating
(AEH)
\citep{Schlegel:81a,Providakes:88,Stauning:89,St.-Maurice:90a,Williams:1992,
Foster:Simultaneous00,Bahcivan:Plasma2007}, largely because the
turbulent electrostatic field,
$\delta\vec{E}=-\nabla\delta\Phi$, has a small component
parallel to the geomagnetic field $\vec{B}_0$
\citep{StMaLaher:85,Providakes:88,DimantMilikh:JGR03,MilikhDimant:JGR03,
Bahcivan:Parallel2006}.
Anomalous electron heating directly affects the
temperature-dependent electron-neutral collision frequency and,
hence, the electron part of the Pedersen conductivity. This
part, however, is usually small compared to the dominant
electron Hall and ion Pedersen conductivities. At the same
time, AEH causes a gradual elevation of the mean plasma density
within the anomalously heated regions through the thermal
reduction of the plasma recombination rate
\citep{Gurevich:78,St.-Maurice:90b,DimantMilikh:JGR03,MilikhGoncharenko:Anom2006}.
The AEH-induced plasma density elevations increase all
conductivities in proportion by as much as a factor of two.
This mechanism requires tens of seconds or even minutes because
of the slow development of the ionization-recombination
equilibrium. If $\vec{E}_0$ changes faster than the
characteristic recombination timescale then its time-averaged
effect on density will be smoothed over field variations.

Also, low-frequency turbulence in the compressible ionospheric
plasma can directly modify local ionospheric conductivities via
a wave-induced non-linear current (NC) associated with plasma
density irregularities
\citep{RogisterJamin:1975,Oppenheim:Evidence97,Buchert:Effect2006}.
The physical nature of NC has been explained in the companion
paper \citep{Dimant:Magnetosphere2011_Budget}. While the rms
turbulent field $\langle\delta\vec{E}^{2}\rangle^{1/2}$ is
comparable to $E_{0}$ (the angular brackets denote spatial and
temporal averaging), the NC is proportional to the density
perturbations that, in saturated FB turbulence, may reach tens
percent at most. As a result, the total NC, considered as a
plasma response to the external electric field $\vec{E}_{0}$,
amounts to only a fraction of the regular electrojet Hall
current. However, in most of the electrojet the NC is directed
largely parallel to $\vec{E}_{0}$, so that it will increase
significantly the smaller Pedersen conductivity. This is
critically important because it is the Pedersen conductivity
that allows the field-aligned currents to close and dissipate
energy. The combined effect of the NC and AEH makes the
ionosphere less resistive. These anomalous conductances may, at
least partially, account for systematic overestimates of the
total cross-polar cap potential in global MHD models that
employ laminar conductivities.

In this paper, using expressions for the NC obtained in
\citet{Dimant:Magnetosphere2011_Budget}, we quantify a feedback
of developed \emph{E}-region turbulence on the global behavior
of the magnetosphere by calculating the corresponding
modifications of local conductivities. In Sect.~\ref{Anomalous
conductivity}, we do this in general terms of given spectra of
density irregularities. In Sect.~\ref{Heuristic Model of
Turbulence}, invoking the appendix, we employ a heuristic model
of non-linearly saturated turbulence \citep{DimantMilikh:JGR03}
which allows us to easily estimate the turbulent
conductivities. The anomalous heating-induced effect manifests
itself via the mean density variations so that it contributes
proportionally into both the laminar and turbulent
conductivities. In Sect.~\ref{Turbulent Frictional Heating in
Global Modeling}, we calculate the turbulent frictional heating
sources for possible inclusion into global
ionosphere-thermosphere models. In Sect.~\ref{Discussion}, we
demonstrate that all anomalous effects combined may nearly
double the undisturbed Pedersen conductance and hence they
should be taken into account when applying inner boundary
conditions in global MHD codes employed for space weather
modeling. In the appendix, using a novel and compact formalism,
we derive the general fluid-model dispersion relation for waves
in arbitrarily magnetized linearly unstable plasma.

\section{Turbulent Ionospheric Conductivity:
\allowbreak Quasilinear Approximation\label{Anomalous
conductivity}}

In this section, we calculate
the turbulent ionospheric conductivity associated with the
\textit{E}-region non-linear current, using the general
quasilinear expressions for the NC obtained in the companion
paper \citep{Dimant:Magnetosphere2011_Budget}. This quasilinear
approach is justified by the relatively small plasma density
perturbations, even though electrostatic field modulations can
be comparable to the driving electric field
\citep{DimantMilikh:JGR03,Oppenheim:Fully2011}.

The turbulent conductivity tensor can be constructed similar to
the laminar conductivity tensor,
\begin{equation}
\overleftrightarrow{\sigma^{\mathrm{L}}}\equiv\left[
\begin{array}
[c]{ccc}%
\sigma_{\mathrm{P}}^{\mathrm{L}} & \sigma_{\mathrm{H}}^{\mathrm{L}} & 0\\
-\sigma_{\mathrm{H}}^{\mathrm{L}} & \sigma_{\mathrm{P}}^{\mathrm{L}} & 0\\
0 & 0 & \sigma_{\parallel}^{0}%
\end{array}
\right]  , \label{sigma_tensor}%
\end{equation}
where the general parallel, Pedersen, and Hall conductivities in a Cartesian
system $x_{1,2,3}$ with the axis $\hat{x}_{3}$ directed along $\vec{B}$ are
given by \citep[e.g.,][]{Kelley:Ionosphere2009}
\begin{subequations}
\label{sigma_II,P,H}%
\begin{align}
\sigma_{\parallel}^{\mathrm{L}}  &  \equiv\frac{\vec{j}_{L\parallel}\cdot\vec
{E}_{0\parallel}}{E_{0\parallel}^{2}}=\frac{(\kappa_{e}+\kappa_{i})ne}%
{B},\label{sigma_II}\\
\sigma_{\mathrm{P}}^{\mathrm{L}}  &  \equiv\frac{\vec{j}_{L\perp}\cdot\vec{E}_{0\perp}
}{E_{\perp}^{2}}=\frac{(\kappa_{e}+\kappa_{i})(1+\kappa_{i}\kappa_{e}%
)ne}{(1+\kappa_{e}^{2})(1+\kappa_{i}^{2})B},\label{sigma_P}\\
\sigma_{\mathrm{H}}^{\mathrm{L}}  &  \equiv\frac{\vec{j}_{L\perp}\cdot(\vec{E}%
_{0}\times\hat{b})}{E_{0\perp}^{2}}=-\ \frac{\left(  \kappa_{e}^{2}-\kappa
_{i}^{2}\right)  ne}{(1+\kappa_{e}^{2})(1+\kappa_{i}^{2})B}, \label{sigma_H}%
\end{align}
\end{subequations}
respectively. Here $n$ is the density of the quasi-neutral plasma;
$\vec{j}_{L\perp,\parallel}$
are the perpendicular and parallel to $\vec{B}_0$ components of the laminar
current density $\vec{j}_{L}$; $\kappa_s\equiv \Omega_s/\nu_{s}$ are the
magnetization parameters for particles of the $s$-kind ($s=e,i$) with
$\Omega_s=eB/m_s$ and $\nu_{sn}$ being the corresponding gyro-frequency and
mean collision frequency with neutrals; $e$ is the (positive) elementary charge;
$m_s$ are the particle-$s$ masses. These conductivities neglect Coulomb collisions,
as compared to electron and ion collisions with neutrals, so that they are
largely applicable to altitudes below the \textit{F}-region ionosphere.

Similarly to Eq.~(\ref{sigma_tensor}), the turbulent
conductivity tensor determined by the non-linear current,
$\vec{j}^{\mathrm{NC}}\equiv\overleftrightarrow{\sigma^{\mathrm{NC}}}%
\cdot\vec{E}_{0}$, can be expressed as
\begin{equation}
\overleftrightarrow{\sigma^{\mathrm{NC}}}\equiv\left[
\begin{array}
[c]{ccc}%
\sigma_{\mathrm{P}}^{\mathrm{NC}} & \sigma_{\mathrm{H}}^{\mathrm{NC}} & 0\\
-\sigma_{\mathrm{H}}^{\mathrm{NC}} & \sigma_{\mathrm{P}}^{\mathrm{NC}} & 0\\
0 & 0 & \sigma_{\parallel}^{\mathrm{NC}}%
\end{array}
\right]\!.  \label{sigma_NC_tensor_3D}%
\end{equation}
We can find $\overleftrightarrow{\sigma^{\mathrm{NC}}}$ using
quasilinear Eq.~(42) from
\citet{Dimant:Magnetosphere2011_Budget} obtained for
arbitrarily magnetized particles,
\begin{align}
&  \vec{j}^{\mathrm{NC}}=\frac{en_{0}}{\kappa_{i}\kappa_{e}}\nonumber\sum_{\vec{k},
\omega\neq0}(\vec{k}\cdot\vec{U}_{0})\left\vert \frac{\delta n_{\vec{k},\omega}}
{n_{0}}\right\vert^{2} \\
&  \times\frac{(1+\kappa_{e}^{2})\left(
1+\kappa_{i}^{2}\right)  \vec{k}_{\parallel}+\left(1+\kappa_{i}\kappa
_{e}\right)  \vec{k}_{\perp}-\left(  \kappa_{e}-\kappa_{i}\right)  (\vec
{k}\times\hat{b})}{(1+\psi_{\vec{k}
})k_{\perp}^{2}}.\label{j_NL}%
\end{align}
This yields
\begin{subequations}
\label{sigma_NC_II,P,H}%
\begin{align}
  \sigma_{\parallel}^{\mathrm{NC}} & =\vec{j}_{\parallel}^{\mathrm{NC}}\cdot
\vec{E}_{0\parallel}/E_{0\parallel}^{2}=
\frac{en_{0}(1+\kappa_{e}^{2})\left(  1+\kappa_{i}^{2}\right)  }
{\kappa_{i}\kappa_{e}}\nonumber\\
& \times \sum_{\vec{k},\omega\neq0}\frac{(\vec{k}_{\parallel
}\cdot\vec{E}_{0\parallel})(\vec{k}\cdot\vec{U}_{0})}{(1+\psi_{\vec{k}%
})k_{\perp}^{2}E_{0\parallel}^{2}}\left\vert \frac{\delta n_{\vec{k},\omega}
}{n_{0}}\right\vert ^{2},\label{sigma_NC_II}\\
\sigma_{\mathrm{P}}^{\mathrm{NC}} &  =\vec{j}_{\perp}^{\mathrm{NC}}\cdot\vec
{E}_{0\perp}/E_{0\perp}^{2}
=\frac{en_{0}}{\kappa_{i}\kappa_{e}}\sum_{\vec{k},\omega\neq0}
(\vec{k}\cdot\vec{U}_{0})\left\vert \frac{\delta n_{\vec{k},\omega}}{n_{0}
}\right\vert ^{2}\nonumber\\
&\times\frac{\left(\kappa_{e}-\kappa_{i}\right)
\vec{k}\cdot(\vec{E}_{0} \times\hat{b})+\left(
1+\kappa_{i}\kappa_{e}\right)  \vec{k}_{\perp}\cdot
\vec{E}_{0}}{(1+\psi_{\vec{k}})k_{\perp}
^{2}E_{0\perp}^{2}},\label{sigma_NC_P}\\
\sigma_{\mathrm{H}}^{\mathrm{NC}} &  =\vec{j}_{\perp}^{\mathrm{NC}}\cdot\left(
\vec{E}_{0}\times\hat{b}\right)  /E_{0\perp}^{2}
  =\frac{en_{0}}{\kappa_{i}\kappa_{e}}\sum_{\vec{k},\omega\neq0}
(\vec{k}\cdot\vec{U}_{0})\left\vert \frac{\delta n_{\vec{k},\omega}}{n_{0}
}\right\vert ^{2}\nonumber\\
&\times\frac{\left(  1+\kappa_{i}\kappa_{e}\right)  \vec{k}\cdot(\vec{E}_{0}
\times\hat{b})-\left(  \kappa_{e}-\kappa_{i}\right)  \vec{k}_{\perp}\cdot
\vec{E}_{0}}{(1+\psi_{\vec{k}})k_{\perp}
^{2}E_{0\perp}^{2}},\label{sigma_NC_H}
\end{align}
\end{subequations}
where, according to Eq.~(24) from \citet{Dimant:Magnetosphere2011_Budget},
\begin{align}
\vec{U}_{0\parallel} &  =-\ \frac{\left(  \kappa_{e}+\kappa_{i}\right)
\vec{E}_{0\parallel}}{B},\nonumber\\
\vec{U}_{0\perp} &  =\frac{\left(  \kappa_{e}+\kappa_{i}\right)  [\left(
\kappa_{e}-\kappa_{i}\right)  (\vec{E}_{0}\times\hat{b})-\left(  1+\kappa
_{i}\kappa_{e}\right)  \vec{E}_{0\perp}]}{\left(  1+\kappa_{e}^{2}\right)
\left(  1+\kappa_{i}^{2}\right)  B}\label{k.U_0}%
\end{align}
are the parallel and perpendicular to $\vec{B}_0$ components of
the laminar relative velocity between the undisturbed electron
and ion streams; $\vec{U}_0\equiv\vec{V}_{e0}-\vec{V}_{i0}$. In
the quasilinear approach, the total conductivity,
$\overleftrightarrow {\sigma^{\mathrm{tot}}}$, is determined by
merely adding the laminar and turbulent conductivities,
\begin{equation}
\overleftrightarrow{\sigma^{\mathrm{tot}}}=\overleftrightarrow{\sigma^{\mathrm{L}}%
}+\overleftrightarrow{\sigma^{\mathrm{NC}}},%
\label{sigma_tot}%
\end{equation}
where each of them is proportional to the same plasma density
$n_0$. Density increases caused by the AEH-induced reduction of
the recombination rate are automatically included if one allows
for the corresponding temperature modifications of
$n_0(T_{e})$.

Global MHD codes employed for magnetosphere modeling assume
equipotential magnetic field lines,
$\vec{E}_{0\parallel}=0$,
across the ionosphere due to high
electron mobility along $\vec{B}_{0}$. In the \textit{E/D}
regions, this approximation is valid  above the $80$\,km
altitude  where $\kappa_{e}\gg1,\kappa_{i}$, i.e., almost in
the entire electrojet. Hence, only conductivities perpendicular
to $\vec{B}_{0}$ matter, while the parallel currents can be
determined using the continuity of the total current density,
$\nabla_{\parallel}\cdot\vec{j}_{\parallel}=-\nabla_{\perp
}\cdot\vec{j}_{\perp}$. In this approach, the high parallel
electron conductivity is not invoked unless one needs to
estimate the tiny parallel field, $E_{0\|}$.

Integration of the quasi-neutral current conservation equation,
$\nabla\cdot\vec{j}^{\mathrm{tot}}\equiv-\nabla\cdot(\overleftrightarrow
{\sigma^{\mathrm{tot}}}\cdot\nabla\Phi_{0})=0$, along nearly
vertical equipotential magnetic field lines yields a 2-D
second-order differential relation for the potential,
$\nabla_{\perp}\left(\overleftrightarrow
{\Sigma^{\mathrm{tot}}}\nabla_{\perp}\Phi_{0}\right)=j_{\|}$,
where $\overleftrightarrow
{\Sigma^{\mathrm{tot}}}=\overleftrightarrow{\Sigma^{\mathrm{L}}}+\overleftrightarrow
{\Sigma^{\mathrm{NC}}}$  is the total height-integrated
ionospheric conductance tensor and $j_{\|}$ is the parallel
current density on top of the conducting ionosphere. This
height-integrated Ohm's law serves as an approximate inner
boundary condition for the global MHD codes
\citep[e.g.,][]{Merkin:Global2005}.

With neglect of $\vec{E}_{0\parallel}$,
$\vec{E}_{0\perp}\approx\vec{E}_{0}=-\nabla\Phi_{0}$, while
assuming $\kappa_{e}\gg1\gtrsim\kappa_{i}$, the normal
conductivity tensor given by Eq.~(\ref{sigma_II,P,H}) reduces
to
\begin{subequations}
\label{sigma_II,P,H_reduced}%
\begin{align}
\sigma_{\mathrm{P}}^{\mathrm{L}}  &  \approx\frac{(1+\kappa_{i}\kappa_{e})ne}%
{\kappa_{e}(1+\kappa_{i}^{2})B}=\frac{\kappa_{i}(1+\psi_{\perp})ne}%
{(1+\kappa_{i}^{2})B},\label{sigma_P_reduced}\\
\sigma_{\mathrm{H}}^{\mathrm{L}}  &  \approx-\ \frac{ne}{(1+\kappa_{i}^{2})B}.
\label{sigma_H_reduced}%
\end{align}
\end{subequations}
Under the same conditions, while additionally presuming
field-aligned irregularities, $k_{\parallel}\ll
k_{\perp}\approx k$, Eqs.~(\ref{sigma_NC_II,P,H}) and
(\ref{k.U_0}) simplify to
\begin{subequations}
\label{sigma_red_NC_P,H}%
\begin{align}
\sigma_{\mathrm{P}}^{\mathrm{NC}}  &  \approx-\ \frac{en_{0}}{\kappa_{i}
\kappa_{e}}\sum_{\vec{k},\omega\neq0}(\vec{k}\cdot\vec{U}_{0})\left\vert
\frac{\delta n_{\vec{k},\omega}}{n_{0}}\right\vert ^{2}\nonumber\\
&\times\frac{\kappa_{e}\vec{k}\cdot(\vec{E}
_{0}\times\hat{b})+\left(  1+\kappa_{i}\kappa_{e}\right)  \vec{k}\cdot\vec
{E}_{0}}{(1+\psi_{\vec{k}})k^{2}E_{0}^{2}},\label{sigma_red_NC_P}
\\
\sigma_{\mathrm{H}}^{\mathrm{NC}}  &  \approx-\ \frac{en_{0}}{\kappa_{i}%
\kappa_{e}}\sum_{\vec{k},\omega\neq0}(\vec{k}\cdot\vec{U}_{0})\left\vert
\frac{\delta n_{\vec{k},\omega}}{n_{0}}\right\vert ^{2}\nonumber\\
&\times\frac{\left(1+\kappa_{i}\kappa
_{e}\right)  \vec{k}\cdot(\vec{E}_{0}\times\hat{b})-\kappa_{e}(\vec{k}%
\cdot\vec{E}_{0})}{(1+\psi_{\vec{k}})k^{2}E_{0}^{2}
},
\label{sigma_red_NC_H}%
\end{align}%
\end{subequations}
\begin{equation}
\vec{U}_{0}=\vec{U}_{0\perp}\approx\frac{\kappa_{e}(\vec{E}_{0}\times\hat
{b})-\left(  1+\kappa_{i}\kappa_{e}\right)  \vec{E}_{0}]}{\kappa_{e}\left(
1+\kappa_{i}^{2}\right)  B}\simeq\frac{\vec{E}_{0}\times\hat{b}-\kappa_{i}%
\vec{E}_{0}}{\left(  1+\kappa_{i}^{2}\right)  B}, \label{k>U_0_red}%
\end{equation}
where the last approximation of Eq.~(\ref{k>U_0_red})
represents the simplest interpolation between lower E/D-region
altitudes with $\kappa_{i}\ll1$, $\kappa_{e} \gg
1\gtrsim\kappa_{i}\kappa_{e}$ and higher \textit{E}-region
altitudes with $\kappa_{i} \sim1$, $\kappa_{i}\kappa_{e}\gg1$.
Quasilinear Eq.~(\ref{sigma_red_NC_P,H}) applies to arbitrary
spectra of \textit{E}-region turbulence, provided the density
perturbations are reasonably small (see the discussion below
related to Fig.~\ref{Fig:Delta n/n_0}). In the next section, we
employ a specific model of non-linearly saturated turbulence to
calculate $\sigma_{\mathrm{P,H}}^{\mathrm{NC}}$.

\section{Turbulent Ionospheric Conductivity: \allowbreak
Heuristic Model
of Turbulence\label{Heuristic Model of Turbulence}}

The current state of  \textit{E}-region instability theory does
not give us accurate spectra of density irregularities $\delta
n_{\vec{k},\omega}$ as functions of the external electric field
and ionospheric parameters. In order to estimate the turbulent
conductivities we are forced to use simplified models of
non-linearly saturated turbulence. Here we employ a heuristic
model of turbulence (HMT) developed previously for qualitative
explanation of AEH by \citet{DimantMilikh:JGR03}. The HMT
provides approximate rms values of the coupled turbulent
electric field and density irregularities. This analytical
model based on simple physical reasoning has been successfully
tested by quantitative comparisons with AEH observations
\citep{MilikhDimant:JGR03,MilikhGoncharenko:Anom2006} and our
recent 3-D PIC simulations \citep{Oppenheim:Fully2011}. A
significant advantage of employing the HMT is that this model
enables one to estimate the turbulent conductivities in terms
of the rms density irregularities, without any knowledge of
their detailed $\vec{k},\omega$-spectrum.

The HMT consists of two major heuristic assumptions of
developed turbulence during the non-linearly saturated stage of
the FB instability \citep{DimantMilikh:JGR03}. First, the model
assumes the effective values of the major perpendicular
component of the turbulent electric field,
$\delta\vec{E}_{\vec{k}\perp}\perp\vec{B}$. Second, it assumes
the effective aspect angles determined by effective
$k_{\parallel }/k_{\perp}$. Both these assumptions refer to a
modified FB-instability threshold field within the developed
turbulence, where the combined linear growth/damping rate
$\gamma_{\vec{k}}$ equals zero. This modified threshold field,
hereinafter referred to as merely the threshold field, is
determined by the same expression as the actual threshold of
instability excitation where the plasma temperatures are
replaced by the elevated temperatures due to anomalous heating.
We determine this threshold field using general expressions for
the Farley-Buneman linear growth rate obtained in the appendix.

To calculate the threshold electric field necessary to maintain
turbulence, $E_{\mathrm{Thr}}^{\min}$, we can write
$\gamma_{\vec{k}}$ in the limit of $\kappa_e\gg 1 \gtrsim
\kappa_i$ as
\begin{equation}
\gamma_{\vec{k}}\approx\frac{\psi_{\vec{k}}k_{\perp}^{2}}{(1+\psi_{\vec{k}%
})\nu_{i}}\left[  \frac{\left(  1-\kappa_{i}^{2}\right)  U_{0}^{2}\cos^{2}%
\chi_{\vec{k}}}{(1+\psi_{\vec{k}})^{2}}-C_{s}^{2}\right]  ,
\label{gamma_total_reduced}%
\end{equation}
where $\chi_{\vec{k}}$ is  the angle between $\vec{k}$ and
$\vec{U}_{0}$ \citep{DimOppen2004:ionthermal1},
$C_{s}=(T_{e}+T_{i})^{1/2}/m_{i}$ is the isothermal
ion-acoustic speed, $U_{0}\equiv|\vec{U}_{0}|$,
\begin{subequations}
\label{psi}%
\begin{align}
\psi_{\vec{k}}  &  \equiv\psi_{\perp}\left[  1+(1+\kappa_{e}^{2})(1+\kappa
_{i}^{2})\frac{k_{\parallel}^{2}}{k_{\perp}^{2}}\right]  ,\label{psi_k}\\
\psi_{\perp}  &  \equiv\frac{1}{\kappa_{i}\kappa_{e}}=\frac{\nu_{e}\nu_{i}%
}{\Omega_{e}\Omega_{i}}, \label{psi_perp}%
\end{align}
\end{subequations}
and we presume a strict inequality of $\kappa_{i}<1$. Using the relation
\begin{equation}
U_{0}\simeq\frac{E_{0}}{(1+\kappa_{i}^{2})^{1/2}B}, \label{U_0_via_E_0}%
\end{equation}
following from Eq.~(\ref{k>U_0_red}), we can rewrite
Eq.~(\ref{gamma_total_reduced}) as%
\begin{equation}
\gamma_{\vec{k}}=\frac{\psi_{\vec{k}}(1-\kappa_{i}^{2})k_{\perp}^{2}}%
{(1+\psi_{\vec{k}})(1+\kappa_{i}^{2})B^{2}\nu_{i}}\left[  \frac{E_{0}^{2}%
\cos^{2}\chi_{\vec{k}}}{(1+\psi_{\vec{k}})^{2}}-\frac{(E_{\mathrm{Thr}}^{\min
})^{2}}{(1+\psi_{\perp})^{2}}\right]  , \label{gamma_total_reduced_2}%
\end{equation}
where $E_{\mathrm{Thr}}^{\min}$ is the minimum FB-instability threshold
electric field at a given altitude for optimally directed waves with $\vec
{k}_{\perp}\parallel\vec{U}_{0}$ and $k_{\parallel}=0$ (i.e., for $\chi
_{\vec{k}}=0$ and $\psi_{\vec{k}}=\psi_{\perp}$),%
\begin{equation}
E_{\mathrm{Thr}}^{\min}=(1+\psi_{\perp})\left(  \frac{1+\kappa_{i}^{2}%
}{1-\kappa_{i}^{2}}\right)  ^{1/2}E_{\mathrm{Thr}}^{(0)}.%
\label{E_Thr}
\end{equation}
Here%
\begin{equation}
E_{\mathrm{Thr}}^{(0)}=C_{s}B\approx20\left(  \frac{T_{e}+T_{i}}%
{600\mathrm{K}}\right)  ^{1/2}\left(  \frac{B}{5\times10^{4}\mathrm{nT}%
}\right)  \mathrm{mV/m}%
\label{E_Thr_0}
\end{equation}
is the absolute threshold-field minimum which can only be
reached at optimum altitudes for the FB instability excitation
where the conditions of $\kappa_{i}^{2}\ll1$ and
$\psi_{\perp}=\Theta_{0}^{2}/\kappa_{i}^{2}\ll1$ overlap (at
high latitudes, this occurs at 100--105~km altitude
\citep[][Fig.~2]{DimOppen2004:ionthermal1};
$\Theta_{0}^{2}\equiv m_{e}\nu_{e}/(m_{i}\nu
_{i})\simeq1.8\times10^{-4}$). Be advised that notations in
this paper slightly differ from those in \citet[][Eqs. (14),
(15) and others]{DimantMilikh:JGR03}. Namely, $E_{0}$ in this
paper corresponds to $E_{C}$ in \citet{DimantMilikh:JGR03},
while our $E_{\mathrm{Thr}}^{(0)}$ corresponds to $E_{0}$ and
$E_{\mathrm{Thr}}^{\min}$ corresponds to $E_{\mathrm{Thr}}$ in
that paper.

The first heuristic assumption specifies the rms perpendicular
turbulent electric field,
\begin{equation}
\langle\delta E_{\perp}^{2}\rangle\simeq\alpha_{1}(E_{0}-E_{\mathrm{Thr}
}^{\min})^{2}, %
\label{E_perp_heuristic}
\end{equation}
where $\alpha_{1}$ is a dimensionless factor of order unity
\citep[][Eq.~(25)]{DimantMilikh:JGR03}. The logic behind
Eq.~(\ref{E_perp_heuristic}) is that in the non-linearly
saturated state the major electron non-linearity
$\propto(\delta\vec{E}_{\perp }\times\delta n)$ balances, on
average, the linear instability growth for optimally directed
waves. For $E_{0}\gg E_{\mathrm{Thr}}^{\min}$, the rms
turbulent field, $\langle\delta E_{\perp}^{2}\rangle^{1/2}$, is
similar in magnitude to $E_{0}$, while near the minimum FB\
instability threshold, $E_{0}\approx E_{\mathrm{Thr}}^{\min}$,
it reduces linearly with $(E_{0}-E_{\mathrm{Thr}}^{\min})$.
This heuristic assumption makes no distinction between the 2-D
and 3-D cases, implying that the rms perpendicular turbulent
fields in both cases are approximately the same. This has been
confirmed by our recent fully kinetic simulations
\citep{Oppenheim:Fully2011}.

The other heuristic assumption distinguishes the 3-D case from
the purely 2-D one by quantifying the effective magnitude of
$k_{\parallel}$. In \textit{E}-region turbulence,
$k_{\parallel}$ is always much less than $k_{\perp}\approx k$,
but even small $k_{\parallel}$ matter because for highly mobile
electrons with $\kappa_e\gg 1$ they significantly modify the
parameter $\psi_{\vec{k}}$ defined by Eq.~(\ref{psi_k}) and,
hence, the linear instability threshold. Also, it is the
parallel turbulent field, $\delta E_\|\propto k_\|$, that
largely cause AEH. The idea of the second HMT assumption is
that the effective values of $k_{\parallel}$ in developed
turbulence, on average, settle the system on the margin of
linear stability where $\gamma_{\vec{k}}=0$. Given the optimum
value of the flow angle for the pure FB instability, the
marginal linear stability yields
\begin{equation}
\psi_{\vec{k}}\simeq\psi_{\vec{k}}^{\mathrm{m}}\equiv\frac{E_{0}%
}{E_{\mathrm{Thr}}^{\min}}(1+\psi_{\perp})-1 \label{1+psi_m}%
\end{equation}
(cf. \citet[][Eq.~(26)]{DimantMilikh:JGR03}  with our
$\psi_{\vec{k}}^{\mathrm{m}}$ corresponding to
$(1+\kappa_{i}^{2})\psi_{\max}$ in \citet{DimantMilikh:JGR03}
(the conditions in the two papers look formally different, but
they are essentially the same because of the negligible
difference $\sim\psi_{\perp}\kappa_{i}^{2}=\Theta_{0}^{2}
\simeq1.8\times10^{-4}$). Equations (27)--(29) from
\citet{DimantMilikh:JGR03} yield specific values for the
effective rms values of the parallel turbulent field in terms
of a second unknown constant of order unity, $\alpha_{2}$. That
was important for determining the average heating source
responsible for AEH but is not required for our current
purposes. Equations~(\ref{E_perp_heuristic}) and
(\ref{1+psi_m}) is all we need from HMT to quantify the rms
density perturbations and hence the corresponding non-linear
current.

In the quasilinear approximation, the turbulent electric field spectrum,
$\delta\vec{E}_{\vec{k},\omega}=-i\vec{k}\delta\Phi_{\vec{k},\omega}$, is
proportional to the density irregularity spectrum $\delta n_{\vec{k},\omega}$.
In our current limit of $\kappa_{e}\gg1>\kappa_{i}$,
Eq.~(28) from \citet{Dimant:Magnetosphere2011_Budget} yields
\begin{equation}
\delta\vec{E}_{\vec{k},\omega} =
\frac{(1+\kappa_{i}^{2})BU_{0}
\cos{\chi_{\vec{k}}}}{\kappa_{i}(1+\psi_{\vec{k}})}\left(  \frac{\delta
n_{\vec{k},\omega}}{n_{0}}\right) , %
\label{delta_E_versus_delta_n}
\end{equation}
where the proportionality coefficient between
$\delta\vec{E}_{\vec{k},\omega}$ and $\delta
n_{\vec{k},\omega}$ depends on the wavevector direction via
$\cos{\chi_{\vec{k}}}=\vec{k}\cdot\vec{U}_{0}/(kU_0)$ and
$\psi_{\vec{k}}$, but is independent of its absolute value,
$k$. This fact allows us to directly relate the rms value of
the entire turbulent field, $\langle\delta E_{\perp}^{2}
\rangle^{1/2}\equiv(\sum_{\vec{k},\omega\neq0}|\delta\vec{E}_{\vec{k},\omega
}|^{2})^{1/2}$, to the rms of the total density fluctuations,
$\langle\delta
n^{2}\rangle^{1/2}\equiv(\sum_{\vec{k},\omega\neq0}|\delta
n_{\vec{k},\omega }|^{2})^{1/2}$. Indeed, if we assume in
accord with simulations
\citep[e.g.,][]{OppenDim2004:ionthermal2,Oppenheim:Large-Scale2008,Oppenheim:Fully2011}
that most of the $\vec{k}_{\perp}$-spectrum is concentrated
within a narrow angular sector around a preferred wavevector
direction
\begin{equation}
\left(\frac{\vec{k}}{k}\right)^{\mathrm{pr}}\simeq\frac{\vec{U}_{0}\cos\chi_{\vec
{k}}^{\mathrm{pr}}+\vec{U}_{0}\times\vec{b}\ \sin\chi_{\vec{k}}^{\mathrm{pr}%
}} {U_{0}} , \label{k_pref}%
\end{equation}
where the flow angle $\chi_{\vec{k}}^{\mathrm{pr}}$ with the
positive sign of $\sin\chi_{\vec{k}}^{\mathrm{pr}}$ corresponds
to the tilt in the $(-\vec {E}_{0})$-direction
\citep{DimOppen2004:ionthermal1}, then, setting for the entire
spectrum $\vec{k}/k\simeq (\vec{k}/k)^{\mathrm{pr}}$,
$\psi_{\vec{k}} \simeq\psi_{\vec{k}}^{\mathrm{m}}$ and using
Eq.~(\ref{U_0_via_E_0}), we obtain from
Eq.~(\ref{delta_E_versus_delta_n}) an approximate relation
\begin{align}
\frac{\langle\delta n^{2}\rangle^{1/2}}{n_{0}}  &  \simeq\frac{\kappa
_{i}(1+\psi_{\vec{k}}^{\mathrm{m}})\langle\delta E_{\perp}^{2}\rangle
^{1/2}\cos\chi_{\vec{k}}^{\mathrm{pr}}}{(1+\kappa_{i}^{2})BU_{0}}\nonumber\\
&  \simeq\frac{\kappa_{i}\alpha_{1}^{1/2}(1+\psi_{\perp})\cos\chi_{\vec{k}%
}^{\mathrm{pr}}}{(1+\kappa_{i}^{2})^{1/2}}\left(  \frac{E_{0}}{E_{\mathrm{Thr}%
}^{\min}}-1\right)  . \label{|del_n|_vs_|del_E|}%
\end{align}
According to Eq.~(\ref{E_Thr}), we have
\begin{equation}
\frac{E_{0}}{E_{\mathrm{Thr}}^{\min}}=\frac{E_{0}}{(1+\psi_{\perp
})E_{\mathrm{Thr}}^{(0)}}\left(  \frac{1-\kappa_{i}^{2}}{1+\kappa_{i}^{2}%
}\right)  ^{1/2} \label{E_0/E_Thr}%
\end{equation}
where $E_{\mathrm{Thr}}^{(0)}$ is defined by Eq.~(\ref{E_Thr_0}).%

\begin{figure}
\noindent\includegraphics[width=21pc]{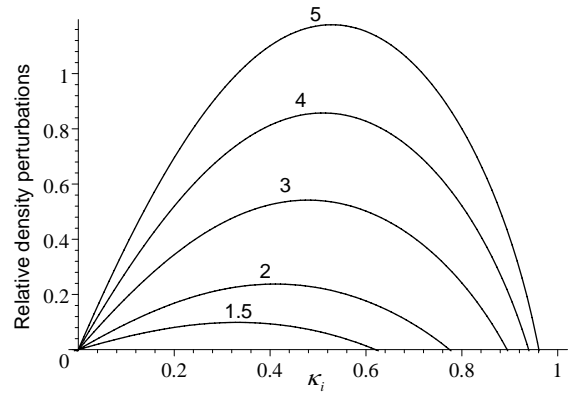}
\caption{Relative density perturbations, $\langle\delta n^2\rangle^{1/2}/n_0$,
vs. the ion magnetization parameter $\kappa_i\equiv \Omega_i/\nu_i$ for different
values of $E_0/E_{\mathrm{Thr}}^{\min}$ (shown near the curves).}
\label{Fig:Delta n/n_0}
\end{figure}

Figure~\ref{Fig:Delta n/n_0}\ based on
Eqs.~(\ref{|del_n|_vs_|del_E|}) and (\ref{E_0/E_Thr}) with
$\alpha_{1}=1$ and $\chi_{\vec{k}}^{\mathrm{pr}}=0$
\citep{DimantMilikh:JGR03} shows that if
$E_{0}/E_{\mathrm{Thr}}^{(0)}$ is large enough then relative
rms density perturbations may even exceed unity. Such
non-physical occasions break the validity of our quasilinear
approximation. We should bear in mind, however, that large
values of $E_{0}$ automatically lead to strong electron and ion
temperature elevations due to combined regular and anomalous
heating. This raises
$E_{\mathrm{Thr}}^{(0)}\propto(T_{e}+T_{i})^{1/2}$ and prevents
$E_{0}/E_{\mathrm{Thr}}^{(0)}$ from reaching too large values
(we discuss this issue in more detail in the next section). For
example, in the extreme case of $E_{0}=160\,$mV/m, the electron
temperature reaches above $4000$\thinspace K
\citep{Bahcivan:Plasma2007} at an altitude around 110~km
corresponding to $\kappa_{i}\simeq 0.3$
\citep[][Fig.~2]{DimOppen2004:ionthermal1}, not counting the
simultaneously increasing ion temperature. According to
Eq.~(\ref{E_Thr_0}), this raises $E_{\mathrm{Thr}}^{(0)}$ above
$50\,$mV/m, making $E_{0}/E_{\mathrm{Thr}}^{(0)}<3$. We believe
that this extreme value of $E_{0}/E_{\mathrm{Thr} }^{(0)}$
imposes the top restriction on possible density perturbations,
keeping them below $50\%$. Less extreme but more typical
density perturbations are expected to be largely within the
reliable limits of the quasilinear approach.

Applying $\vec{k}/\vec{k}\simeq
(\vec{k}/\vec{k})^{\mathrm{pr}}$, while replacing $(1+\psi
_{\vec{k}}^{\mathrm{m}})$ according to Eq.~(\ref{1+psi_m}) and
$\langle\delta n^{2}\rangle^{1/2}$ according to
Eq.~(\ref{|del_n|_vs_|del_E|}), we use
Eqs.~(\ref{sigma_red_NC_P,H}) and (\ref{k>U_0_red}) to express
the turbulent conductivities in terms of the corresponding
laminar conductivities given by
Eq.~(\ref{sigma_II,P,H_reduced}) as
\begin{subequations}
\label{sigma_final_heuristic}%
\begin{align}
\sigma_{\mathrm{P}}^{\mathrm{NC}}  &  \simeq\frac{\alpha_{1}\left[  \left(
1-\kappa_{i}^{2}\right)  \cos\chi_{\vec{k}}-2\kappa_{i}\sin\chi_{\vec{k}%
}\right]  \cos^{3}\chi_{\vec{k}}}{1+\kappa_{i}^{2}}\nonumber\\
&  \times\left(  \frac{E_{0}}{E_{\mathrm{Thr}}^{\min}}-1\right)  \left(
1-\frac{E_{\mathrm{Thr}}^{\min}}{E_{0}}\right)  \sigma_{\mathrm{P}}
^{\mathrm{L}},\label{sigma_final_heuristic_P}\\
\sigma_{\mathrm{H}}^{\mathrm{NC}}  &  \simeq-\ \frac{\alpha_{1}\left[
2\kappa_{i}\cos\chi_{\vec{k}}+\left(  1-\kappa_{i}^{2}\right)  \sin\chi
_{\vec{k}}\right]  (1+\psi_{\perp})\cos^{3}\chi_{\vec{k}}}{1+\kappa_{i}^{2}%
}\nonumber\\
&  \times\left(  \frac{E_{0}}{E_{\mathrm{Thr}}^{\min}}-1\right)  \left(
1-\frac{E_{\mathrm{Thr}}^{\min}}{E_{0}}\right)  \sigma_{\mathrm{H}}^{\mathrm{L}}.
\label{sigma_final_heuristic_H}%
\end{align}
\end{subequations}
These relations are only applicable for $E_{0}\geq
E_{\mathrm{Thr}}^{\min}$ and $\kappa_i\leq 1$, otherwise
$\sigma_{\mathrm{P,H}}^{\mathrm{NC}}=0$. Furthermore, according
to Eqs.~(\ref{E_Thr}) and (\ref{E_Thr_0}), the threshold field
$E_{\mathrm{Thr}}^{\min}\propto E_{\mathrm{Thr}}^{(0)}\propto
(T_e + T_i)^{1/2}$ depends strongly on the electron and ion
temperatures. This means that the self-consistent inclusion of
the turbulent conductivity in the total conductivity tensor
requires simultaneously including AEH
\citep{DimantMilikh:JGR03,MilikhDimant:JGR03}, otherwise the
effect of NC-induced turbulent conductivity will be exaggerated
dramatically.

The AEH-induced increase in the plasma density due to the
temperature reduction of the recombination rate
\citep{DimantMilikh:JGR03,MilikhGoncharenko:Anom2006} will
raise all values in the conductivity tensor in proportion to
the increased plasma density. This slowly-developing effect,
however, can slightly decrease if the strong external electric
field $\vec{E}_0$ varies too fast
\citep{Codrescu:Importance1995,Codrescu:Electric2000,Matsuo:Effects2008,
Cosgrove:Comparison2009,
CosgroveCodrescu:Electric2009}.

\section{Turbulent Frictional Heating in Global Modeling
\label{Turbulent Frictional Heating in Global Modeling}}

An accurate description of strong electric-field perturbations,
such as storms, substorms, and sub-auroral polar streams,
requires including macroscopic effects of \textit{E}-region
turbulence into global computer models, like the Coupled
Magnetosphere Ionosphere Thermosphere (CMIT) model
\citep{Wiltberger:Initial2004,Wang:Initial2004}. This model has
been created by combining the magnetosphere MHD solver LFM
\citep{LyonFedder:Lyon-Fedder-Mobarry2004} and
ionosphere-thermosphere NCAR solver TIEGCM
\citep{RobleRidley:Thermosphere1994,Wang:High-resolution1999}.
The ring-current model RCM \citep{Toffoletto:Inner2003Review}
has been added recently, as well as TIEGCM has been extended to
cover the mesosphere (TIMEGCM). Currently, the CMIT-RCM model
includes ionospheric processes associated only with laminar
conductivities. The turbulence-induced non-linear current (NC)
can be incorporated by adding turbulent conductivities given by
Eq~(\ref{sigma_final_heuristic}). Self-consistency requires
including also anomalous plasma heating caused by turbulent
fields. The AEH increases (through the reduced recombination
rate) the plasma density and, hence, all conductivities in
proportion. However, the anomalous temperature elevations exert
a negative feedback on the saturated level of the plasma
turbulence, reducing the NC and its non-linear Pedersen
conductivity.

Effects of \textit{E}-region turbulence can be included
directly in TIMEGCM and other codes that that model
high-latitude neutral and plasma dynamics and chemical
reactions. Given accurate heating sources, TIMEGCM includes all
ionization/recombination and collisional cooling processes
automatically. To account for temperature inputs caused by both
laminar and turbulent fields, the energy balance equation for
each species should include the corresponding source of
frictional heating. Currently, TIMEGCM includes the laminar ion
and neutral frictional heating but neglects the electron one.
This is a reasonable approximation for altitudes above 100 km
under quiet ionospheric conditions. However, during strong
electric-field events, the \textit{E}-region turbulence via AEH
dramatically raises the electron temperature, meaning that the
total electron frictional heating should also be included. For
the convection field well above the instability threshold field
given by Eqs.~(\ref{sigma_final_heuristic}), ion turbulent
heating can also be appreciable (for $E_{0}\gg E_{\mathrm{Thr}
}^{\min}$, it is comparable to the ion laminar heating).
Accuracy of the entire global model requires including the
corresponding neutral frictional heating as well. Note that
highly anisotropic \textit{E}-region turbulence also causes
average momentum changes in the plasma. Plasma-neutral
collisions in turn can transfer these changes to neutral
particles. For the weakly ionized \textit{E} region, however,
relative momentum changes are much less important than relative
energy changes, so that we will ignore the former and focus
entirely on the latter.

In the multi-component \textit{E}-region ionosphere where
electron-neutral (\emph{e-n}) and ion-neutral (\emph{i-n})
collisions dominate, we can calculate the total laminar and
turbulent frictional heating of electrons,
$H_{e}^{\mathrm{tot}}$, ions, $H_{i}^{\mathrm{tot}}$, and
neutrals, $H_{n}^{\mathrm{tot}}$, using a quasilinear
approximation to quadratic accuracy. In terms of the average
turbulent perturbations, these are given by
\begin{align}
H_{e}^{\mathrm{tot}}  &  =m_{e}\nu_{e}\left[  n_{0}\left(  V_{e0}%
^{2}+\left\langle \delta V_{e}^{2}\right\rangle \right)  +2(\vec{V}_{e0}%
\cdot\langle\delta n\delta\vec{V}_{e}\rangle)\right]  ,\nonumber\\
H_{i}^{\mathrm{tot}}  &  =\sum_{n}\frac{m_{i}m_{n}\nu_{in}}{m_{i}+m_{n}%
}\nonumber\\
&  \times\left[  n_{i0}\left(  V_{i0}^{2}+\left\langle \delta V_{i}%
^{2}\right\rangle \right)  +2(\vec{V}_{i0}\cdot\langle\delta n_{i}\delta
\vec{V}_{i}\rangle)\right]  ,\label{Heatingi_all}\\
H_{n}^{\mathrm{tot}}  &  =\sum_{i}\frac{m_{i}^{2}\nu_{in}}{m_{i}+m_{n}%
}\nonumber\\
&  \times\left[  n_{i0}\left(  V_{i0}^{2}+\left\langle \delta V_{i}%
^{2}\right\rangle \right)  +2(\vec{V}_{i0}\cdot\langle\delta n_{i}\delta
\vec{V}_{i}\rangle)\right]  ,\nonumber
\end{align}
with the summations taken over all possible collisions between ions and neutrals,
where indices $i$ and $n$ refer to separate groups of ions or neutrals; $\sum
_{i}n_{i0}=n_{0}$. The physical meaning of various terms in the square
brackets is explained in \citet{Dimant:Magnetosphere2011_Budget}. The
expressions for $H_{i,n}^{\mathrm{tot}}$ explicitly
take into account the fact that only $m_{n}/(m_{i}+m_{n})$ of the partial
\emph{i-n} electric field heating, $m_{i}\nu_{in}V_{i}^{2}$, goes to ions
($H_{i}^{\mathrm{tot}}$), while the remaining fraction goes directly to the
colliding neutrals ($H_{n}^{\mathrm{tot}}$). The total electron collision
frequency $\nu_{e}$ includes all electron-neutral (\emph{e-n}) collisions,
$\nu_{e}=\sum_{n}\nu_{en}$; but because $m_{e}\ll m_{n}$, the
contributions to $H_{n}^{\mathrm{tot}}$ from the \emph{e-n} collisions are
negligible. Equation~(\ref{Heatingi_all}) is written in the neutral frame of
reference, presuming that all thermosphere components move with a common
neutral-wind velocity.

In a multi-fluid plasma, the laminar and turbulent velocities,
$\vec{V}_{s0}$ and $\delta\vec{V}_{s}$, can be expressed in
terms of the convection field $\vec {E}_{0}$ and spectral
harmonics of the wave potential $\delta\Phi_{\vec {k},\omega}$
similarly to the two-fluid plasma
\citep{Dimant:Magnetosphere2011_Budget}. However, the
first-order linear relations between the harmonics of the ion
spectral densities, $\delta n_{i\vec{k},\omega}$, and
potential, $\delta\Phi_{\vec{k},\omega}$, which are crucial for
obtaining equations like Eq.~(\ref{delta_E_versus_delta_n}),
(\ref{delta_Phi_via_delta_n_final}), can be rather complicated
or cannot even be expressed in any closed analytical form. The
reason is that the algebraic order of the underlying linear
dispersion relation for the first-order wave frequency,
$\omega_{\vec{k}}$, in the general case equals the number of
separate ion species. Even for the two dominant \emph{E}-region
ion species, NO$^{+}$ and O$_{2}^{+}$, when the dispersion
relation reduces to a quadratic equation, the first-order
relations become cumbersome. These relations, however, simplify
dramatically if all ion species have a common magnetization
parameter, $\kappa_{i}=\Omega_{i}/\nu_{i}$
($\nu_{i}\equiv\sum_{n}\nu_{in}$), so that all ion species
respond to the fields equally. As a result, the multi-fluid
relation between $\delta\Phi_{\vec {k},\omega}$ and $\delta
n_{\vec{k},\omega}$ and various terms in $n_{s0}\left(
V_{s0}^{2}+\left\langle \delta V_{s}^{2}\right\rangle \right)
+2(\vec{V}_{s0}\cdot\langle\delta
n_{s}\delta\vec{V}_{s}\rangle)$ ($s=e,i$) reduce to two-fluid
Eqs.~(49) and (51) from
\citet{Dimant:Magnetosphere2011_Budget}. For
$\kappa_{e}\gg1\gtrsim\kappa_{i}$, these become
\begin{equation}
\nu_{e}n_{0}V_{e0}^{2}\approx\frac{\nu_{e}n_{0}E_{0}^{2}}{B^{2}},\qquad\nu
_{i}n_{i0}V_{i0}^{2}=\frac{\nu_{i}n_{i0}\kappa_{i}^{2}E_{0}^{2}}{\left(
1+\kappa_{i}^{2}\right)  B^{2}},\label{zero-order-heat}%
\end{equation}%
\begin{subequations}
\label{diag_psi}%
\begin{align}
\nu_{e}n_{0}\left\langle \delta V_{e}^{2}\right\rangle  &  \approx\frac
{m_{i}\nu_{i}n_{0}(1+\kappa_{i}^{2})}{m_{e}}\sum_{\vec{k},\omega\neq0}%
\frac{\psi_{\vec{k}}(\vec{k}\cdot\vec{U}_{0})^{2}}{(1+\psi_{\vec{k}}%
)^{2}k_{\perp}^{2}}\left\vert \frac{\delta n_{\vec{k},\omega}}{n_{0}%
}\right\vert ^{2},\label{diag_psi_ee}\\
\nu_{i}n_{i0}\left\langle \delta V_{i}^{2}\right\rangle  &  \approx\nu
_{i}n_{i0}(1+\kappa_{i}^{2})\sum_{\vec{k},\omega\neq0}\frac{(\vec{k}\cdot
\vec{U}_{0})^{2}}{(1+\psi_{\vec{k}})^{2}k_{\perp}^{2}}\left\vert \frac{\delta
n_{\vec{k},\omega}}{n_{0}}\right\vert ^{2},\label{diag_psi_ii}%
\end{align}%
\end{subequations}
\begin{subequations}
\label{``diag''_2}%
\begin{align}
2\nu_{e}(\vec{V}_{e0}\cdot\langle\delta n\delta\vec{V}_{e}\rangle) &
\approx\frac{2en_{0}(1+\kappa_{i}^{2})}{m_{e}\kappa_{i}\kappa_{e}}\nonumber\\
&\sum
_{\vec{k},\omega\neq0}\frac{(\vec{k}\cdot\vec{E}_{0})(\vec{k}\cdot\vec{U}
_{0})}{(1+\psi_{\vec{k}})k_{\perp}^{2}}\left\vert \frac{\delta
n_{\vec
{k},\omega}}{n_{0}}\right\vert ^{2},\label{``diag''_2_e}\\
2\nu_{i}(\vec{V}_{i0}\cdot\langle\delta n_{i}\delta\vec{V}_{i}\rangle) &
\approx\frac{2en_{i0}}{m_{i}}\sum_{\vec{k},\omega\neq0}\frac{(\vec{k}\cdot
\vec{E}_{0})(\vec{k}\cdot\vec{U}_{0})}{(1+\psi_{\vec{k}})k_{\perp}^{2}%
}\left\vert \frac{\delta n_{\vec{k},\omega}}{n_{0}}\right\vert ^{2}%
.\label{``diag''_2_i}%
\end{align}
where $\delta n_{\vec{k},\omega}=\sum_{i}\delta
n_{i\vec{k},\omega}$ is the electron density, $\psi_{\vec{k}}$
and $\vec{U}_{0}$ are given by Eqs.~(\ref{psi}) and
(\ref{U_0_via_E_0}), and
$(1+\kappa_{i}^{2})(1+\kappa_{e}^{2}k_{\parallel}^{2}/k_{\perp}^{2})$
emerging in the derivation of Eq.~(\ref{diag_psi_ee}) was
approximated by
$\psi_{\vec{k}}/\psi_{\perp}=\kappa_{e}\kappa_{i}\psi_{\vec{k}}$
\citep[][see Eq.~(34b) and text below
it]{Dimant:Magnetosphere2011_Budget}.

To calculate heating rates in terms of \textit{E}-region
parameters, we need estimates of the turbulent density
perturbations. Using the heuristic model of saturated
turbulence, along with the approach that lead us from
Eq.~(\ref{sigma_red_NC_P,H}) via Eqs.~(\ref{E_perp_heuristic}),
(\ref{1+psi_m}), and (\ref{|del_n|_vs_|del_E|}) to
Eq.~(\ref{sigma_final_heuristic}), we obtain
\end{subequations}
\begin{align}
&\sum_{\vec{k},\omega\neq0}\frac{\psi_{\vec{k}}(\vec{k}\cdot\vec{U}_{0})^{2}%
}{(1+\psi_{\vec{k}})^{2}k_{\perp}^{2}}\left\vert \frac{\delta n_{\vec
{k},\omega}}{n_{0}}\right\vert ^{2}  \simeq\frac{\alpha_{1}\kappa_{i}%
^{2}(E_{0}-E_{\mathrm{Thr}}^{\min})^{2}\cos^{4}\chi_{\vec{k}}^{\mathrm{pr}}%
}{(1+\kappa_{i}^{2})^{3}B^{2}}\nonumber\\
& \times\left[  \frac{E_{0}}{E_{\mathrm{Thr}}^{\min}}\left(  1+\psi_{\perp
}\right)  -1\right],  \nonumber\\
&\sum_{\vec{k},\omega\neq0}\frac{(\vec{k}\cdot\vec{U}_{0})^{2}}{(1+\psi
_{\vec{k}})^{2}k_{\perp}^{2}}\left\vert \frac{\delta n_{\vec{k},\omega}}%
{n_{0}}\right\vert ^{2}   \simeq\frac{\alpha_{1}\kappa_{i}^{2}(E_{0}%
-E_{\mathrm{Thr}}^{\min})^{2}\cos^{4}\chi_{\vec{k}}^{\mathrm{pr}}}%
{(1+\kappa_{i}^{2})^{3}B^{2}}, \nonumber\\
&\sum_{\vec{k},\omega\neq0} \frac{(\vec{k}\cdot\vec{E}_{0})(\vec{k}\cdot\vec
{U}_{0})}{(1+\psi_{\vec{k}})k_{\perp}^{2}}\left\vert \frac{\delta n_{\vec
{k},\omega}}{n_{0}}\right\vert ^{2} \nonumber\\
& \simeq\frac{\alpha_{1}\kappa_{i}%
^{2}(E_{\mathrm{Thr}}^{\min}-E_{0})^{2}\cos^{3}\chi_{\vec{k}}^{\mathrm{pr}%
}\sin\chi_{\vec{k}}^{\mathrm{pr}}}{(1+\kappa_{i}^{2})^{5/2}B}. \label{podstavy}%
\end{align}
Equations (\ref{Heatingi_all})--(\ref{podstavy}) provide
frictional heating terms for the potential inclusion into the
corresponding energy-balance equations of TIMEGCM or similar
ionosphere-thermosphere models.
Equations~(\ref{sigma_final_heuristic})--(\ref{Heatingi_all})
provide a reasonable estimate of  turbulent conductivity and
heating and should greatly improve modeling of the geospace
environment during disturbed conditions.

\section{Discussion\label{Discussion}}

\begin{figure}
\noindent\includegraphics[width=0.48\textwidth]{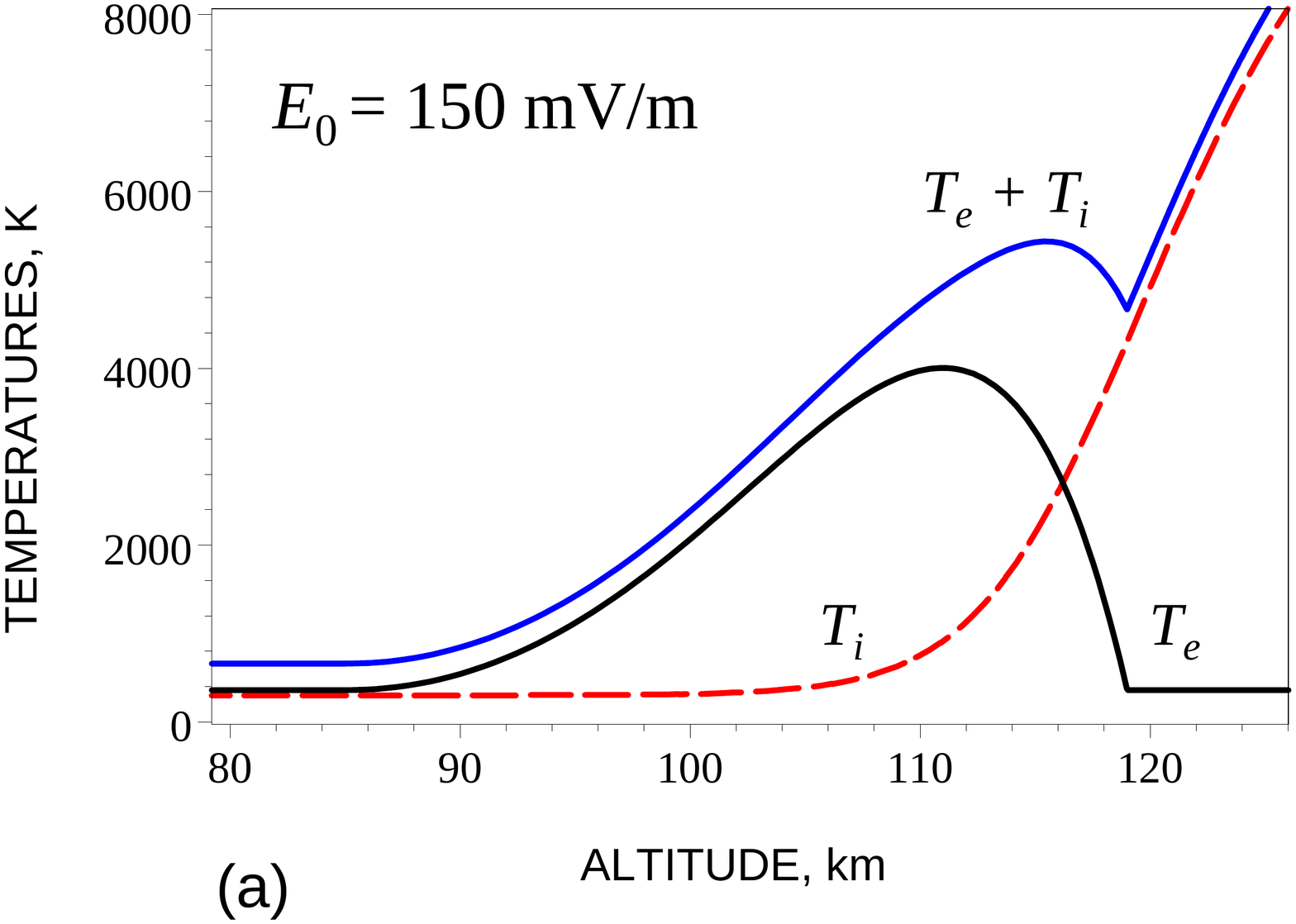}
\hfill
\noindent\includegraphics[width=0.48\textwidth]{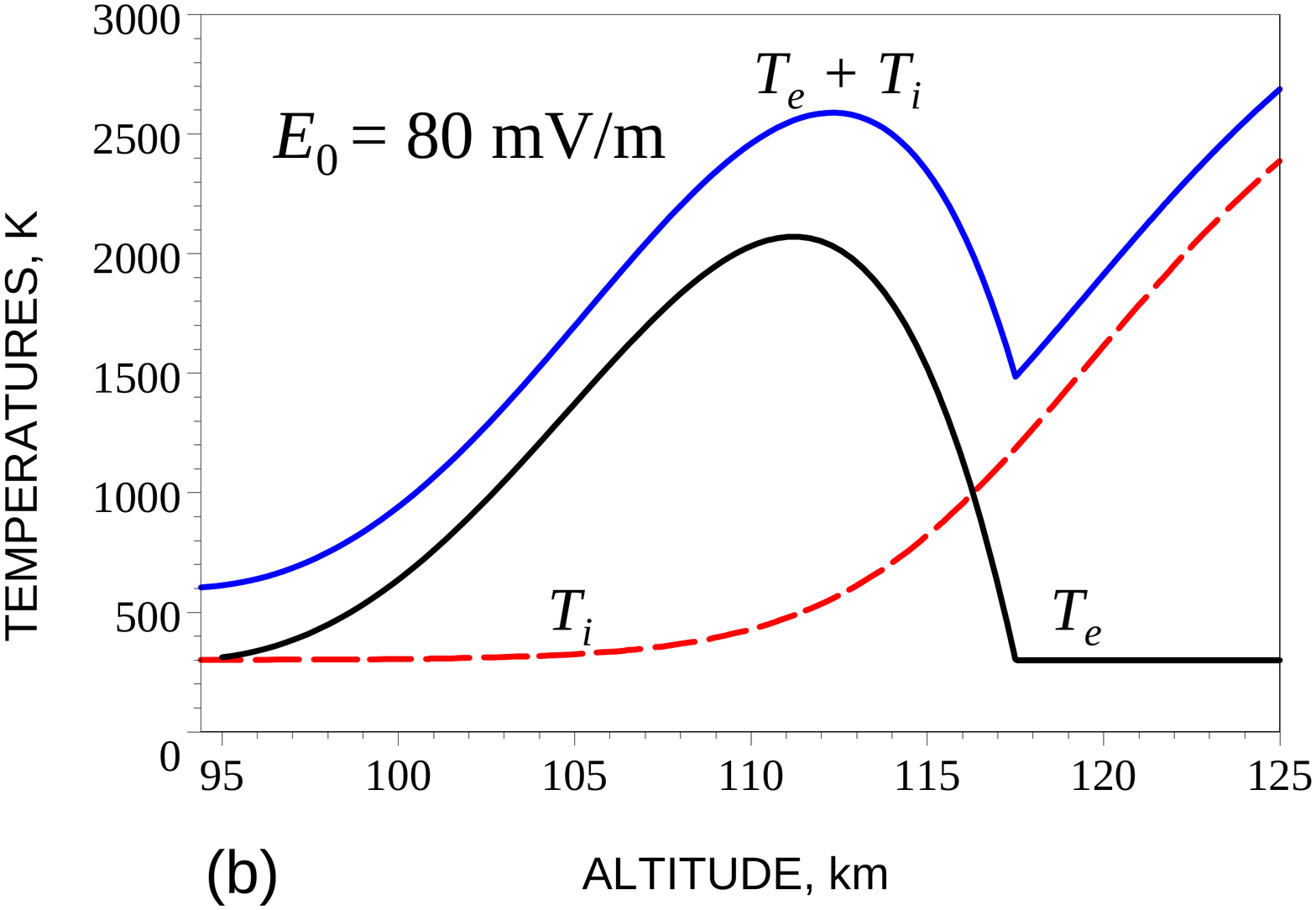}
\caption{\label{Fig:Temper_150} Altitude dependence of the
electron and ion temperatures, $T_e$ and $T_i$, for two
different values of $E_0$ (shown in figures). The electron
temperature includes self-consistent AEH according to
\citet{MilikhDimant:JGR03}. The kinks in $T_e$ and $T_e+T_i$
near the 119.5~km altitude (a) and 117.5~km (b) are caused by
sharp disappearance of instability and, hence, AEH.}
\end{figure}

\begin{figure}
\noindent\includegraphics[width=0.48\textwidth]{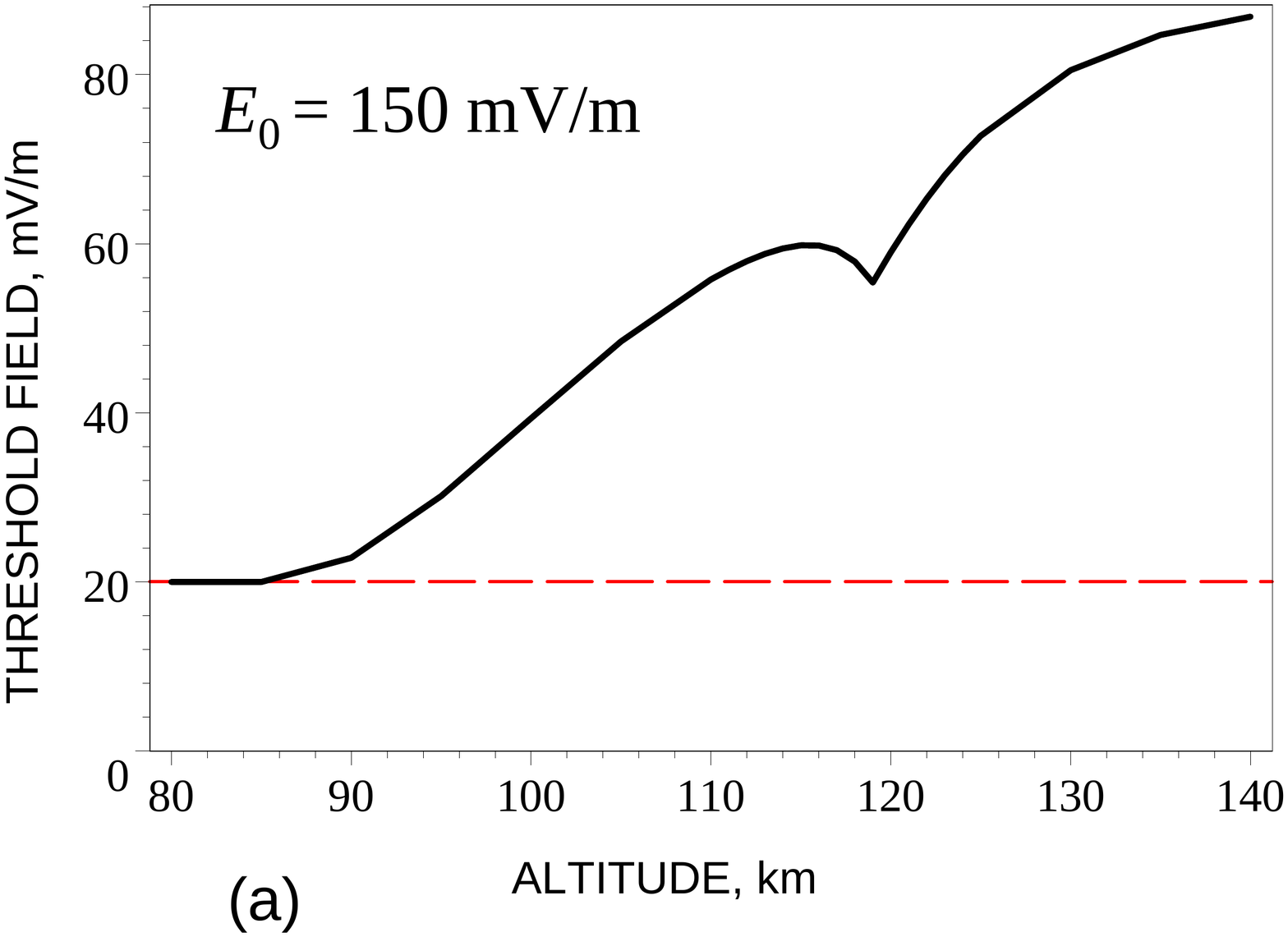} \hfill
\noindent\includegraphics[width=0.48\textwidth]{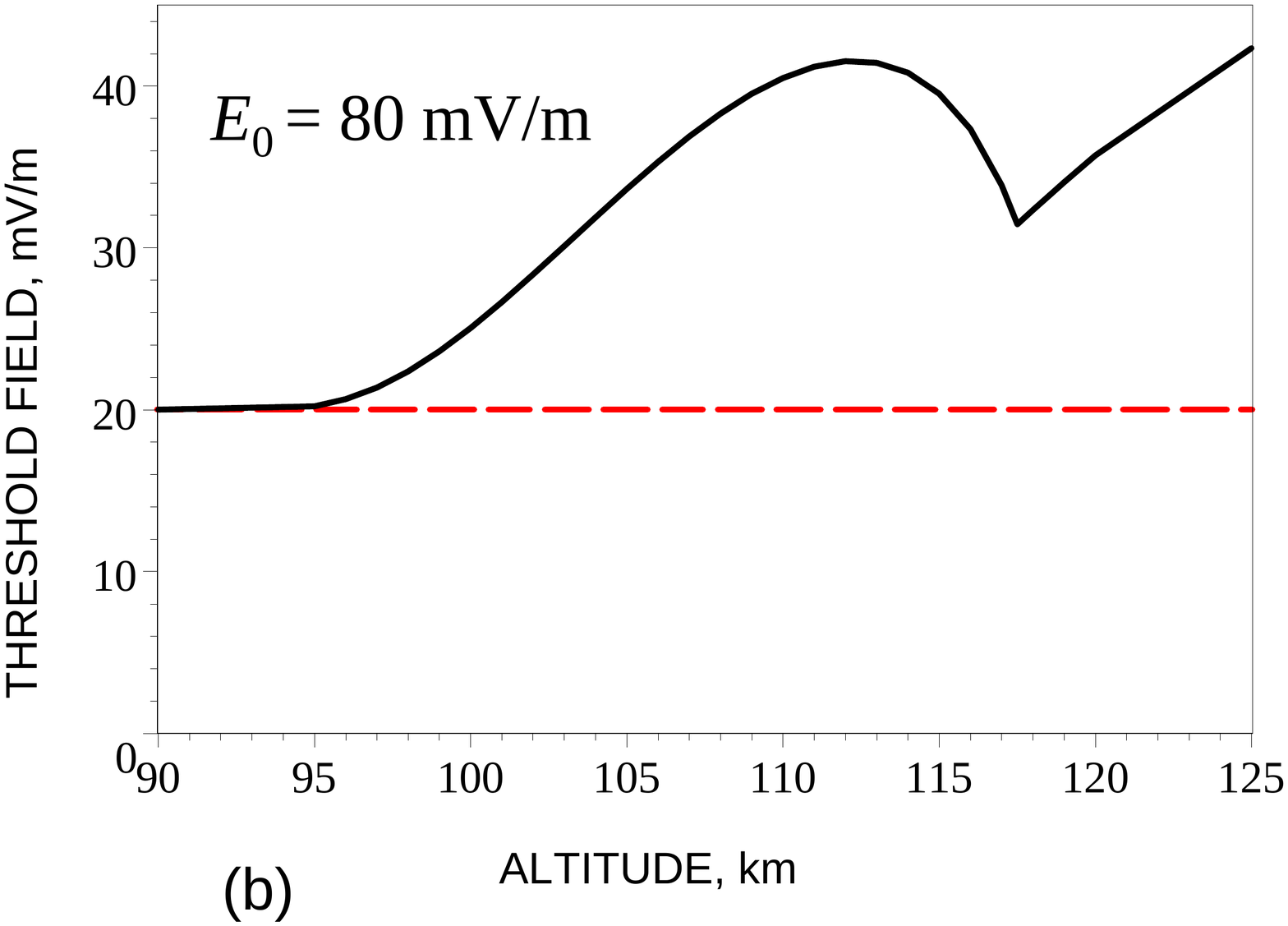}
\caption{Altitude dependence of the FB-instability threshold electric field,
$E_{\mathrm{Thr}}^{\min}$, Eqs.~(\ref{E_Thr}) and (\ref{E_Thr_0}), elevated
due to particle heating for the same values $E_0$ as in Fig.~\ref{Fig:Temper_150}.
The kinks in $E_{\mathrm{Thr}}^{\min}\propto (T_e+T_i)^{1/2}$ correspond to those
in  Fig.~\ref{Fig:Temper_150}.}
\label{Fig:thresh_field_150}
\end{figure}

\begin{figure}
\noindent\includegraphics[width=21pc]{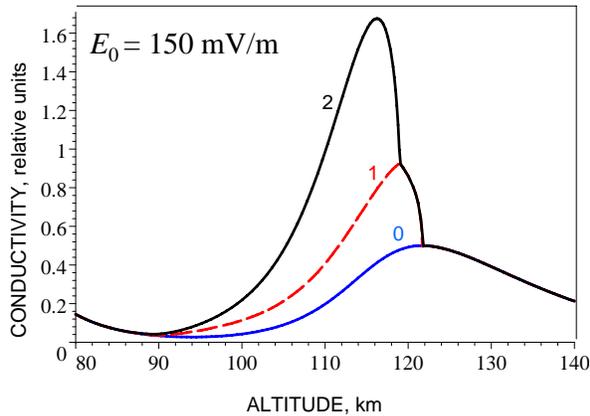}
\caption{Altitude dependence of the total -- normal and anomalous --
Pedersen conductivity for $E_0=150$~mV/m (in relative units). Curve 0
shows the undisturbed conductivity. Curve 1 includes additionally the
NC-induced anomalous conductivity calculated according to
Eq.~(\ref{sigma_final_heuristic_P}) with $\alpha_1=0$ and
$\chi_{\vec{k}}=0$. Curve 2 shows the total Pedersen conductivity
with the AEH-affected plasma density elevated according to the
steady-state ionization recombination model by \citet{MilikhGoncharenko:Anom2006}.
The NC-induced anomalous conductivity disappears above the ion magnetization
boundary $\simeq 122$~km; the AEH-recombination effect vanishes above the top
boundary of anomalous heating $\simeq 119.5$~km (see the corresponding kinks
in Figs.~\ref{Fig:Temper_150} and \ref{Fig:thresh_field_150}).}
\label{Fig:conduct_150}
\end{figure}

\begin{figure}
\noindent\includegraphics[width=21pc]{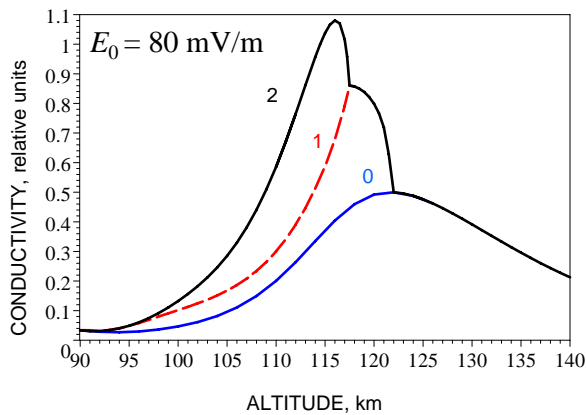}
\caption{The same as Fig.~\ref{Fig:conduct_150}, but for $E_0=80$~mV/m.}
\label{Fig:conduct_80}
\end{figure}

Figures~\ref{Fig:Temper_150}--\ref{Fig:conduct_80} illustrate
anomalous heating and give examples of the anomalously modified
Pedersen conductivity for two cases: an extreme field of
$150$~mV/m and a still large but more modest case of $80$~mV/m.
These figures are based on typical high-latitude ionospheric
parameters and the same values of the major HMT parameter,
$\alpha_1=1$, as in \citet{MilikhDimant:JGR03}.
Figure~\ref{Fig:Temper_150} shows the effect of AEH and normal
ion heating. The combined temperature, $T_e+T_i$, shows kinks
at altitudes where AEH sharply disappears, while ion heating
continues its steep rising. These altitudes are slightly below
the ion magnetization boundary ($\kappa_i=1$) located at
$h_{1}\simeq 122$~km. These kinks translate to those in the
threshold electric field, $E_{\mathrm{Thr}}^{\min} \propto\
(T_e+T_i)^{1/2}$, as seen in Fig.~\ref{Fig:thresh_field_150}.
Figure~\ref{Fig:thresh_field_150} (a) shows that above 100~km
of altitude the threshold field increases dramatically due to
plasma heating, keeping $E_{0}/E_{\mathrm{Thr}}^{\min}$ at a
modest level of 2--3. The much weaker field of $E_{0}=80$~mV/m
(b) gives a slightly lower ratio because the plasma heating in
this case is also much lower. According to Fig.~\ref{Fig:Delta
n/n_0}, such moderate ratio of $E_{0}/E_{\mathrm{Thr}}^{\min}$,
even for the extreme case of $E_0=150$~mV/m, keeps the total
density fluctuation level below 50\%, i.e., roughly within the
framework of the quasilinear approach (see above). This also
reduces the entire effect of NC on the conductivity, as seen
from Fig.~\ref{Fig:conduct_150}.

Figure~\ref{Fig:conduct_150} shows a typical altitudinal
profile of the undisturbed conductivity (curve 0) along with
two curves including the anomalous conductivity. The NC effect
alone (curve 1) nearly doubles the conductivity in a broad
altitude range between 90~km and the ion magnetization boundary
located at an altitude $h_{1}\simeq 122$~km, fairly close to
the maximum of the normal Pedersen conductivity. The
AEH-induced ionization-recombination effect doubles the total
conductivity once again (curve 2), but this occurs in a
slightly narrower range restricted from above by the AEH upper
boundary, $h_{0}\simeq 119.5$~km. Presuming that the lower
altitudes located below the undisturbed conductivity maximum
near $h_{1}$ include about a half of the entire
height-integrated conductance, we see that the NC-induced
anomalous conductivity alone (curve 1) can contribute up to a
half of the undisturbed conductance, while both anomalous
effects combined (curve 2) can nearly double the entire
conductance. We should bear in mind, however, the caveats
associated with slow evolution of ionization-recombination
processes superposed on rapidly varying electric field (see the
discussion above).

Largely the same situation takes place for the smaller field of
$E_0=80$~mV/m (Fig.~\ref{Fig:conduct_80}). Since in this case
the particle-temperature elevation is more modest than that for
$E_0=150$~mV/m, the instability threshold field shown in
Figure~\ref{Fig:thresh_field_150}~(b) remains noticeably lower
than that in Fig.~\ref{Fig:thresh_field_150}~(a). As a result,
the ratio $E_0/E_{\mathrm{Thr}}^{\min}$ changes much less
dramatically compared to $E_0=150$~mV/m. That is why in
Fig.~\ref{Fig:conduct_80} the NC-induced effect for
$E_0=80$~mV/m turns out to be nearly as strong as that in
Fig.~\ref{Fig:conduct_150} for $E_0=150$~mV/m, while the more
fragile AEH-induced recombination effect is much less
pronounced because of lower electron heating.

Above we presumed  a rather conservative value for the major
HMT parameter, $\alpha_1=1$, the same as in
\citet{DimantMilikh:JGR03,MilikhDimant:JGR03,Merkin:Anomalous2005,
MilikhGoncharenko:Anom2006}. Recent 2D and 3D PIC simulations
\citep{Oppenheim:Fully2011} suggest that $\alpha_{1}$ can
actually be closer to $1.5$. This would make
$\sigma_{\mathrm{P}}^{\mathrm{NC}}/\sigma_{\mathrm{P}}^{\mathrm{L}}$
roughly 50\% larger, although due to larger $\langle\delta
n^{2}\rangle^{1/2}/n_{0}$ one may have stronger restrictions on
the quasilinear approach. We should also note that our current
heuristic model of developed turbulence does not include  the
ion thermal driving mechanism (ITM)
\citep{DimOppen2004:ionthermal1,OppenDim2004:ionthermal2}.

\onecolumn

\twocolumn

That is why all our anomalous effects occur strictly below the
ion magnetization boundary, $\kappa_{i}=1$. Possible inclusion
of the ITM would play a two-fold role. On the one hand, this
would expand the altitudinal range of anomalous conductivity to
at least a few kilometers higher
\citep{DimOppen2004:ionthermal1} and increase the total
ionospheric conductance accordingly. A modest reduction in
$\kappa_{i}$ due to anomalously heated ions might also help. On
the other hand, near the ion magnetization boundary the
preferred values of the modified flow angle $\chi_{\vec{k}}$
deviate from $\chi_{\vec{k}}=0$, also due to the ITM as
explained in \citet{DimOppen2004:ionthermal1}. This deviation
of $\chi_{\vec{k}}$ would slightly reduce the
$\chi_{\vec{k}},\kappa_{i}$-dependent factor in the RHS of
Eq.~(\ref{sigma_final_heuristic_P}) and hence lead to a smaller
Pedersen conductivity compared to that for $\chi_{\vec{k}}=0$.
An accurate inclusion of the ITM to the heuristic model of
saturated turbulence, however, would complicate the entire
model. We plan to improve the treatment of the top electrojet
altitudes in future by using both analytical theory and PIC
simulations. At this moment, a better accuracy is of less
importance than the mere fact that the instability-induced
conductivity occupies roughly the entire lower half of the
Pedersen conductive layer and can nearly double the whole
conductance. Note also that polarized sporadic-\textit{E}
clouds \citep{Dimant:Interaction2010}, or even ubiquitous
meteor trails \citep{Dimant:Meteor:External2009}, can also make
additional contributions to anomalous conductances. All these
effects combined can, at least partially, explain why global
MHD codes developed for predictive modeling of space weather
which use normal conductances often overestimate the
cross-polar cap potentials by close to a factor of two.

\section{Summary and Conclusions}

Plasma turbulence generated by \textit{E}-region instabilities
significantly modifies the ionospheric Pedersen conductance by
exciting a net non-linear current,
$\vec{j}^{\mathrm{NC}}=\overleftrightarrow
{\sigma_{\mathrm{P}}^{\mathrm{NC}}}\cdot\vec{E}_{0}$. The
magnitude of $\vec {j}^{\mathrm{NC}}$ can only be a fraction of
the undisturbed Hall current, but can be comparable in
magnitude to the Pedersen current and pointing in that
direction. This creates an additional turbulent conductivity.
Also, anomalous electron heating, via reduced plasma
recombination,  increases the mean plasma density $n_0$ and
all, laminar and turbulent, conductances in proportion with
$n_0$. Estimates based on a quasilinear theory and heuristic
model of non-linearly saturated plasma turbulence show that the
anomalous effects combined can nearly double the Pedersen
conductance, as predicted by Eqs.~(\ref{sigma_red_NC_P,H}) and
(\ref{sigma_final_heuristic}) and illustrated by
Figs.~\ref{Fig:conduct_150} and \ref{Fig:conduct_80}. Such
ionospheric response to the magnetospheric field $\vec{E}_{0}$
may efficiently reduce the high-latitude ionospheric resistance
and decrease the cross-polar cap potential. This effect might
explain, at least partially, why routine MHD simulations with
the normal ionospheric conductances systematically overestimate
this potential by a factor of two. The anomalous effects on the
conductances, as well as turbulent frictional heating, should
be included in global MHD codes developed for space weather
prediction.

\section*{Appendix: General FB/GD Dispersion Relation for
Arbitrarily Magnetized Plasmas}

This appendix develops the two-fluid linear theory of the
Farley-Buneman (FB) and gradient drift (GD) instabilities for
arbitrary magnetization parameters,
$\kappa_{e,i}\equiv\Omega_{e,i}/\nu_{e,i}$, that vary from
small values at lower altitudes to large values at higher
altitudes. This analysis has never been published but is
relevant to the lower ionosphere and enables one to accurately
estimate conventionally neglected terms that may matter in some
cases.

The approach described below allows one to obtain the
fluid-model dispersion relation in a reasonably compact and
physically clear way. Compared to Sect.~3.1 of
\citet{Dimant:Magnetosphere2011_Budget}, here we additionally
include the particle pressure, inertia, regular gradients in
the background plasma density, and recombination. These factors
are responsible for the FB and GD instability drivers, as well
as for the stabilizing diffusion and ionization balance. For
simplicity, we do not include non-isothermal processes
responsible for driving of thermal instabilities
\citep{Dimant:Physical97,Kagan:thermal00,DimOppen2004:ionthermal1}
and ignore the stabilizing effect of a weak non-quasineutrality
\citep{RosenbergChow:98,Kovalev:Modeling2008}. These factors
can be added using essentially the same approach.

The quasineutral two-fluid continuity equations including
ionization and recombination balance are
\begin{equation}
\partial_{t}n+\nabla\cdot(n\vec{V}_{e})=\partial_{t}n+\nabla\cdot(n\vec{V}%
_{i})=Q-\alpha n^{2}, \label{contiki}%
\end{equation}
where $Q$ is the total source of ionization and $\alpha$ is the recombination
constant. Equation~(\ref{contiki}) includes the standard quasineutral relation
$\nabla\cdot(n\vec{U})=0$, where $\vec{U}\equiv\vec{V}_{e}-\vec{V}_{i}$.

The continuity equations for a stationary ($\partial_t n_0 =0$) but inhomogeneous
background density, $n_{0}$, with the corresponding undisturbed fluid
velocities, $\vec{V}_{e,i0}$, are
\begin{equation}
\nabla\cdot(n_{0}\vec{V}_{e0})=\nabla\cdot(n_{0}\vec{V}_{i0})=Q-\alpha
_{0}n_{0}^{2}. \label{contiki_zero}%
\end{equation}
From the left equality, the divergence of the relative velocity, $\vec{U}%
_{0}\equiv\vec{V}_{e0}-\vec{V}_{i0}$, can be expressed in terms of the
background density gradient as%
\begin{equation}
\nabla\cdot\vec{U}_{0}=-\vec{U}_{0}\cdot\nabla n_{0}/n_{0}.
\label{div_U_0_via_grad}%
\end{equation}
In what follows, we presume that both $\vec{U}_{0}$ and $n_{0}$
are known functions of spatial coordinates that satisfy
Eq.~(\ref{div_U_0_via_grad}). Then, to express the zero-order
drift velocities of electrons and ions in terms of $\vec{U}_0$,
we can use Eqs.~(26) and (27) from
\citet{Dimant:Magnetosphere2011_Budget} to obtain
\begin{equation}
\vec{V}_{e0\parallel}=\frac{\kappa_{e}\vec{U}_{0\parallel}}{\kappa_{e}%
+\kappa_{i}},\qquad\vec{V}_{i0\parallel}=-\ \frac{\kappa_{i}\vec
{U}_{0\parallel}}{\kappa_{e}+\kappa_{i}}, \label{V_0_II_via_U}%
\end{equation}
\begin{subequations}
\label{V_0_perp_via_U}%
\begin{align}
\vec{V}_{e0\perp}  &  =\frac{\kappa_{e}[\vec{U}_{0\perp}-\kappa_{i}(\vec
{U}_{0}\times\hat{b})]}{\kappa_{e}+\kappa_{i}},\label{V_0_perp_via_U_e}\\
\vec{V}_{i0\perp}  &  =-\ \frac{\kappa_{i}[\vec{U}_{0\perp}+\kappa_{e}(\vec
{U}_{0}\times\hat{b})]}{\kappa_{e}+\kappa_{i}}. \label{V_0_perp_via_U_i}%
\end{align}
\end{subequations}
Note that these expressions are only valid when neglecting for
the background plasma pressure gradients and gravity,
comparably small effects for almost all ionospheric conditions.

We will now consider Fourier harmonics of wave perturbations, $\delta n_{\vec{k}%
}$, $\delta\vec{E}_{\vec{k}}$, $\delta\vec{V}_{e,i\vec{k}}\propto\exp
[i(\vec{k}\cdot\vec{r}-\omega_{\vec{k}}t)]$, where the wavevector $\vec{k}$
and the complex linear wave frequency,
$\omega_{\vec{k}}=\omega_{\vec{k}}^{\prime}+i\gamma_{\vec{k}}$,
are locally defined. Here $\omega_{\vec{k}}^{\prime}$ is the
real wave frequency, while $\gamma_{\vec{k}}\equiv
-i\operatorname{Im}\,\omega_{\vec{k}}$ is the wave total growth
or damping rate.

Linearizing Eq.~(\ref{contiki_zero}) with respect to
$\delta n_{\vec{k}}$, $\delta\vec{V}_{e,i\vec{k}}$ and introducing two shifted
complex wave frequencies,
\begin{subequations}
\label{shifted _Omega_k}%
\begin{align}
\Omega_{\vec{k}}^{(e)}  &  \equiv\omega_{\vec{k}}-\hat{K}\cdot\vec{V}%
_{e0}+2i\alpha n_{0},\label{shifted _Omega_k_e}\\
\Omega_{\vec{k}}^{(i)}  &  \equiv\omega_{\vec{k}}-\hat{K}\cdot\vec{V}%
_{i0}+2i\alpha n_{0}=\Omega_{\vec{k}}^{(e)}+q, \label{shifted _Omega_k_i}%
\end{align}
\end{subequations}
with
\begin{equation}
\hat{K}\equiv\vec{k}-i\nabla,\qquad q\equiv\hat{K}\cdot\vec{U}_{0}, \label{Kq}%
\end{equation}
we obtain%
\begin{equation}
\Omega_{\vec{k}}^{(e)}\ \frac{\delta n_{\vec{k}}}{n_{0}}=\vec{p}\cdot
\delta\vec{V}_{e\vec{k}},\qquad\Omega_{\vec{k}}^{(i)}\ \frac{\delta n_{\vec
{k}}}{n_{0}}=\vec{p}\cdot\delta\vec{V}_{i\vec{k}}, \label{Omegas_prom}%
\end{equation}
where%
\begin{equation}
\vec{p}\equiv\frac{\vec{K}n_{0}}{n_{0}}=\vec{k}-\frac{i\nabla n_{0}}{n_{0}}.
\label{p}%
\end{equation}

Now we express $\delta\vec{V}_{e,i\vec{k}}$ in terms of $\delta
n_{\vec{k}}$ and $\delta\Phi_{\vec{k},\omega}$,
$\delta\vec{E}_{\vec{k}}=-i\vec{k}\delta\Phi_{\vec{k},\omega}$,
using the momentum equations for the individual electron and
ion mean fluid velocities in the neutral frame,
\begin{subequations}
\label{fluid_momentum}%
\begin{align}
m_{e}\ \frac{\mathrm{d}_{e}\vec{V}_{e}}{\mathrm{d}t}  &  =-e(\vec{E}+\vec{V}_{e}
\times\vec
{B})-\frac{\nabla\left(  nT_{e}\right)  }{n}-m_{e}\nu_{e}\vec{V}%
_{e},\label{fluid_momentum_e}\\
m_{i}\ \frac{\mathrm{d}_{i}\vec{V}_{i}}{\mathrm{d}t}  &  =e(\vec{E}+\vec{V}_{i}
\times\vec
{B})-\frac{\nabla\left(  nT_{i}\right)  }{n}-m_{i}\nu_{i}\vec{V}_{i},
\label{fluid_momentum_i}%
\end{align}
\end{subequations}
where
$\mathrm{d}_{e,i}/\mathrm{d}t\equiv\partial_{t}+\vec{V}_{e,i}\cdot\nabla$
are the full derivatives for both electrons and ions. For
low-frequency \emph{E}/\emph{D}-region plasma processes, the
electron inertia described by the LHS\ of
Eq.~(\ref{fluid_momentum_e}), unlike the ion inertia in
Eq.~(\ref{fluid_momentum_i}), is usually neglected, but we will
keep it for completeness and symmetry. Neglecting in
Eq.~(\ref{fluid_momentum}) temperature perturbations, we
express the velocity perturbations as
\begin{subequations}
\label{delta_V_prom}%
\begin{align}
\delta\vec{V}_{e\vec{k}}  &  =\vec{G}_{e}\left(  \delta\Phi_{\vec{k},\omega}%
-\frac{T_{e}}{e}\ \frac{\delta n_{\vec{k}}}{n_{0}}\right)
,\label{delta_V_prom_e}\\
\delta\vec{V}_{i\vec{k}}  &  =\vec{G}_{i}\left(  \delta\Phi_{\vec{k},\omega}%
+\frac{T_{i}}{e}\ \frac{\delta n_{\vec{k}}}{n_{0}}\right)  ,
\label{delta_V_prom_i}%
\end{align}
\end{subequations}
where the vector-functions $\vec{G}_{e,i}$,
\begin{subequations}
\label{G}%
\begin{align}
\vec{G}_{e\parallel}  &  \approx\frac{i\tilde{\kappa}_{e}\vec{k}_{\parallel}%
}{B},\qquad\vec{G}_{e\perp}\approx\frac{i\tilde{\kappa}_{e}(\vec{k}_{\perp
}-\tilde{\kappa}_{e}\vec{k}\times\hat{b})}{\left(  1+\tilde{\kappa}_{e}%
^{2}\right)  B},\label{G_e}\\
\vec{G}_{i\parallel}  &  \approx-\ \frac{i\tilde{\kappa}_{i}\vec{k}%
_{\parallel}}{B},\qquad\vec{G}_{i\perp}\approx-\ \frac{i\tilde{\kappa}%
_{i}(\vec{k}_{\perp}+\tilde{\kappa}_{i}\vec{k}\times\hat{b})}{\left(
1+\tilde{\kappa}_{i}^{2}\right)  B}, \label{G_i}%
\end{align}
\end{subequations}
describe anisotropic two-fluid responses to the harmonic perturbations of the
potential and pressures combined. Here, in accord with Eq.~(\ref{ner-vo_T}),
\begin{subequations}
\label{tilde_kappa}%
\begin{align}
\tilde{\kappa}_{e}  &  \equiv\frac{\Omega_{e}}{\nu_{e}-i\Omega_{\vec{k}}%
^{(e)}}\approx\kappa_{e}\left(  1+\frac{i\Omega_{\vec{k}1}^{(e)}}{\nu_{e}%
}\right)  ,\label{tilde_kappa_e}\\
\tilde{\kappa}_{i}  &  \equiv\frac{\Omega_{i}}{\nu_{i}-i\Omega_{\vec{k}}%
^{(i)}}\approx\kappa_{i}\left(  1+\frac{i\Omega_{\vec{k}1}^{(i)}}{\nu_{i}%
}\right)  \label{tilde_kappa_i}%
\end{align}
\end{subequations}
are modified magnetization ratios that include small contributions from the
particle inertia; $\Omega_{\vec{k}1}^{(e,i)}=\operatorname{Re}\Omega_{\vec{k}%
}^{(e,i)}=\omega_{\vec{k}}-\vec{k}\cdot\vec{V}_{e,i0}$.
The small ion inertia contribution described by
$i\Omega_{\vec{k}1}^{(,i)}/\nu_{i}$ is crucial for excitation
of the FB instability.

Now we substitute the expressions for
$\delta\vec{V}_{e,i\vec{k}}$ from Eq.~(\ref{delta_V_prom}) to
Eq.~(\ref{Omegas_prom}), expressing $\Omega
_{\vec{k}}^{(e)}=\Omega_{\vec{k}}^{(i)}-q$ via
$q=\vec{k}\cdot\vec{U}
_{0}-i\nabla\cdot\vec{U}_{0}=\vec{U}_{0}\cdot(\vec{p})^{\ast}$,
where we used Eq.~(\ref{div_U_0_via_grad}) and (\ref{Kq}). This
yields two independent linear relations between $\delta
n_{\vec{k}}$ and $\delta\Phi_{\vec{k},\omega}$:
\begin{align*}
\left[  \Omega_{\vec{k}}^{(e)}+\frac{T_{e}}{e}(\vec{p}\cdot\vec{G}%
_{e})\right]  \frac{\delta n_{\vec{k}}}{n_{0}}  &  =(\vec{p}\cdot\vec{G}%
_{e})\delta\Phi_{\vec{k},\omega},\\
\left[  \Omega_{\vec{k}}^{(e)}+q-\frac{T_{i}}{e}(\vec{p}\cdot\vec{G}%
_{i})\right]  \frac{\delta n_{\vec{k}}}{n_{0}}  &  =(\vec{p}\cdot\vec{G}%
_{i})\delta\Phi_{\vec{k},\omega}.
\end{align*}
From these relations and Eq.~(\ref{shifted _Omega_k_i}), we
obtain two symmetric expressions for the coupled shifted
frequencies,
\begin{subequations}
\label{Omega_general}%
\begin{align}
\Omega_{\vec{k}}^{(e)}  &  =-\ \frac{(\vec{p}\cdot\vec{G}_{e})[q-(\vec{p}%
\cdot\vec{G}_{i})\left(  T_{e}+T_{i}\right)  /e]}{\vec{p}\cdot(\vec{G}%
_{e}-\vec{G}_{i})},\label{Omega_general_e}\\
\Omega_{\vec{k}}^{(i)}  &  =-\ \frac{(\vec{p}\cdot\vec{G}_{i})[q-(\vec{p}%
\cdot\vec{G}_{e})\left(  T_{e}+T_{i}\right)  /e]}{\vec{p}\cdot(\vec{G}%
_{e}-\vec{G}_{i})}, \label{Omega_general_i}%
\end{align}
\end{subequations}
and the general relation between harmonics $\delta\Phi_{\vec{k},\omega}$ and $\delta
n_{\vec{k}}$,%
\begin{equation}
\delta\Phi_{\vec{k},\omega}=-\ \frac{q-\vec{p}\cdot(\vec{G}_{e}T_{e}+\vec{G}_{i}%
T_{i})/e}{\vec{p}\cdot(\vec{G}_{e}-\vec{G}_{i})}\ \frac{\delta n_{\vec{k}}%
}{n_{0}}. \label{delta_Phi_via_delta_n_final}%
\end{equation}
Using the definitions of $\Omega_{\vec{k}}^{(e)}$ or $\Omega_{\vec{k}}^{(i)}$
given by Eq.~(\ref{shifted _Omega_k}), we obtain the complex wave frequency in the
neutral frame, $\omega_{\vec{k}}=\Omega_{\vec{k}}^{(e)}+\hat{K}\cdot\vec
{V}_{e0}-2i\alpha n_{0}=\Omega_{\vec{k}}^{(i)}+\hat{K}\cdot\vec{V}%
_{i0}-2i\alpha n_{0}$,%
\begin{align}
&  \omega_{\vec{k}}\equiv\omega_{\vec{k}}^{\prime}+i\gamma_{\vec{k}}%
=\frac{(\vec{p}\cdot\vec{G}_{e})(\hat{K}\cdot\vec{V}_{i0})-(\vec{p}\cdot
\vec{G}_{i})(\hat{K}\cdot\vec{V}_{e0})}{\vec{p}\cdot(\vec{G}_{e}-\vec{G}_{i}%
)}\nonumber\\
&  +\frac{(\vec{p}\cdot\vec{G}_{i})(\vec{p}\cdot\vec{G}_{e})(T_{e}+T_{i}%
)}{\vec{p}\cdot(\vec{G}_{e}-\vec{G}_{i})e}-2i\alpha n_{0}.
\label{omega_k_general}%
\end{align}
Equations~(\ref{Omega_general}) to (\ref{omega_k_general}) give
the general expressions for the complex wave frequencies. To
find the combined linear growth of instabilities or wave
dissipation, $\gamma_{\vec{k}}=-i\,\operatorname{Im}
\omega_{\vec{k}}$, they should be further split into the real
and imaginary parts.

The approximate two-fluid description of local quasi-harmonic waves is valid
provided the characteristic wavevectors and frequencies satisfy
\begin{subequations}
\label{ner-vo}%
\begin{align}
& L_{\parallel,\perp}^{-1}   \ll k_{\parallel,\perp}\ll l_{e,i}^{-1}
,\rho_{e,i}^{-1},\label{ner-vo_L}\\
& T^{-1},|\gamma
_{\vec{k}}|    \ll|\omega_{\vec{k}}^{\prime}|\ll\nu_{e,i}. \label{ner-vo_T}
\end{align}
\end{subequations}
Here $l_{e,i}$ are the typical mean free paths of the
corresponding particles with respect to ion-neutral and
electron-neutral collisions, $\rho_{e,i}$ are the gyroradii (if
the corresponding particles are magnetized), and $T$ and
$L_{\parallel,\perp}$ are typical temporal and spatial
(parallel and perpendicular to $\vec{B}_{0}$) scales of
ionospheric density variation. Equation~(\ref{ner-vo_L}) should
be satisfied separately for the parallel and perpendicular
directions. Assuming these conditions, we will treat the wave
pressure gradients, local gradients in the background density,
and particle inertia as second-order effects with respect to
the small parameters defined by Eq.~(\ref{ner-vo}). In this
treatment, the first-order factors discussed in
\citet{Dimant:Magnetosphere2011_Budget} define the real wave
frequency, $\omega_{\vec{k}}^{\prime}$, while all second-order
factors combined define the total linear damping or growth
rate, $\gamma_{\vec{k}}$.

Under conditions specified by Eq.~(\ref{ner-vo}), the dominant
real part of the wave frequency, $\omega_{\vec{k}}^{\prime}$,
is determined by the first-order accuracy,
$\omega_{\vec{k}}^{\prime}\approx\omega_{\vec{k}1}$, when
neglecting the pressure gradients, particle inertia, regular
gradients of the background plasma density, and recombination.
To this accuracy, we have $\hat{K}
\approx\vec{p}\approx\vec{k}$,
$q\approx\vec{k}\cdot\vec{U}_{0}$,
$\tilde{\kappa}_{e,i}=\kappa_{e,i}$, and hence
\begin{subequations}
\label{G_1}%
\begin{align}
\vec{G}_{e\parallel}  &  \approx\vec{G}_{e1\parallel}=\frac{i\kappa_{e}\vec
{k}_{\parallel}}{B},\ \ \ \ \vec{G}_{e\perp}\approx\vec{G}_{e1\perp}%
=\frac{i\kappa_{e}(\vec{k}_{\perp}-\kappa_{e}\vec{k}\times\hat{b})}{\left(
1+\kappa_{e}^{2}\right)  B},\label{G_1_e}\\
\vec{G}_{i\parallel}  &  \approx\vec{G}_{i1\parallel}=-\ \frac{i\kappa_{i}%
\vec{k}_{\parallel}}{B},\ \,\vec{G}_{i\perp}\approx\vec{G}_{i1\perp
}=-\ \frac{i\kappa_{i}(\vec{k}_{\perp}+\kappa_{i}\vec{k}\times\hat{b}%
)}{\left(  1+\kappa_{i}^{2}\right)  B}, \label{G_1_i}%
\end{align}%
\end{subequations}
\begin{align}
\vec{p}\cdot\vec{G}_{e}  &  \approx\vec{k}\cdot\vec{G}_{e1}=\frac{i\kappa
_{e}k_{\perp}^{2}}{(1+\kappa_{e}^{2})B}\left[  1+(1+\kappa_{e}^{2}%
)\frac{k_{\parallel}^{2}}{k_{\perp}^{2}}\right]  ,\nonumber\\
\vec{p}\cdot\vec{G}_{i}  &  \approx\vec{k}\cdot\vec{G}_{i1}=-\ \frac
{i\kappa_{i}k_{\perp}^{2}}{(1+\kappa_{i}^{2})B}\left[  1+(1+\kappa_{i}%
^{2})\frac{k_{\parallel}^{2}}{k_{\perp}^{2}}\right]  . \label{proma}%
\end{align}
Using Eqs.~(\ref{psi}), we obtain
\begin{equation}
\vec{p}\cdot(\vec{G}_{e}-\vec{G}_{i})\approx\vec{k}\cdot(\vec{G}_{e1}-\vec
{G}_{i1})=\frac{i\kappa_{i}\kappa_{e}\left(  \kappa_{e}+\kappa_{i}\right)
\left(  1+\psi_{\vec{k}}\right)  k_{\perp}^{2}}{\left(  1+\kappa_{e}%
^{2}\right)  \left(  1+\kappa_{i}^{2}\right)  B}, \label{proma_2}%
\end{equation}
After neglecting small terms proportional to $\left(  T_{e}+T_{i}\right)  $
and $\alpha n_{0}$, Eq.~(\ref{omega_k_general}) yields%
\[
\omega_{\vec{k}}\approx\omega_{\vec{k}1}=\frac{(\vec{k}\cdot\vec{G}_{e1}%
)(\vec{k}\cdot\vec{V}_{i0})-(\vec{k}\cdot\vec{G}_{i1})(\vec{k}\cdot\vec
{V}_{e0})}{\vec{k}\cdot(\vec{G}_{e1}-\vec{G}_{i1})}%
\]
and reduces to Eq.~(37) from
\citet{Dimant:Magnetosphere2011_Budget} with the corresponding
real shifted frequencies,
$\Omega_{\vec{k}1}^{(e,i)}\equiv\operatorname{Re}\Omega_{\vec{k}
}^{(e,i)}=\omega_{\vec{k}1}-\vec{k}\cdot\vec{V}_{e,i0}$,%
\begin{equation}
\Omega_{\vec{k}1}^{(e)}=-\ \frac{(\vec{k}\cdot\vec{U}_{0})(\vec{k}\cdot\vec
{G}_{e1})}{\vec{k}\cdot(\vec{G}_{e1}-\vec{G}_{i1})},\qquad\Omega_{\vec{k}%
1}^{(i)}=-\ \frac{(\vec{k}\cdot\vec{U}_{0})(\vec{k}\cdot\vec{G}_{i1})}{\vec
{k}\cdot(\vec{G}_{e1}-\vec{G}_{i1})}, \label{Omega_k_1}%
\end{equation}
given explicitly by the companion paper Eqs.~(36) and (39).

Now we will develop the dispersion relationships to the
second-order accuracy by taking into account all previously
neglected factors: the wave pressure gradients, particle
inertia, gradients of the zero-order plasma parameters, and
recombination. All second-order terms in
Eq.~(\ref{omega_k_general}), representing linear corrections to
the first-order real frequency $\omega_{\vec{k}1}$, are purely
imaginary, so that their linear combination determines the
total linear growth/damping rate, $\gamma_{\vec{k}}$.

We start by discussing the two last terms in the RHS of
Eq.~(\ref{omega_k_general}). The second-order term proportional to
$(T_{e}+T_{i})$ originates from perturbations of the particle fluid pressure.
To the leading-order accuracy, the term
\begin{align}
&  \frac{(\vec{p}\cdot\vec{G}_{i})(\vec{p}\cdot\vec{G}_{e})(T_{e}+T_{i})}%
{\vec{p}\cdot(\vec{G}_{e}-\vec{G}_{i})e}\approx\frac{(\vec{k}\cdot\vec{G}%
_{i1})(\vec{k}\cdot\vec{G}_{e1})(T_{e}+T_{i})}{\vec{k}\cdot(\vec{G}_{e1}%
-\vec{G}_{i1})e}\nonumber\\
&  =-\ \frac{ik_{\perp}^{2}\left[  1+(1+\kappa_{e}^{2})k_{\parallel}%
^{2}/k_{\perp}^{2}\right]  \left[  1+(1+\kappa_{i}^{2})k_{\parallel}%
^{2}/k_{\perp}^{2}\right]  (T_{e}+T_{i})}{(\kappa_{e}+\kappa_{i})(1+\psi
_{\vec{k}})eB} \label{T_ei_multiplier}%
\end{align}
contributes directly to $\gamma_{\vec{k}}$. The combination of the last two
terms describes the major linear wave dissipation due to particle diffusion
and recombination. For fully magnetized electrons, $\kappa_{e}\equiv\Omega
_{e}/\nu_{e}\gg1$, and unmagnetized ions, $\kappa_{i}\equiv\Omega_{i}/\nu
_{i}\ll1$, Eq.~(\ref{T_ei_multiplier}) reduces to the conventional loss term,
\[
\frac{(\vec{p}\cdot\vec{G}_{i})(\vec{p}\cdot\vec{G}_{e})(T_{e}+T_{i})}{\vec
{p}\cdot(\vec{G}_{e}-\vec{G}_{i})e}\approx-\ \frac{i\psi_{\vec{k}}k_{\perp
}^{2}C_{s}^{2}}{(1+\psi_{\vec{k}})\nu_{i}},
\]
where in this limit $\psi_{\vec{k}}\approx(1+\kappa_{e}^{2}k_{\parallel}%
^{2}/k_{\perp}^{2})\nu_{e}\nu_{i}/(\Omega_{e}\Omega_{i})$.

The first term in the RHS of Eq.~(\ref{omega_k_general}) includes the major
instability drivers. Unlike the two last terms, to the leading accuracy it is
a first-order term. To retrieve second-order corrections, we have to linearize
it with respect to small perturbations $\propto i\Omega_{\vec{k}%
}^{(e,i)}/\nu_{e,i}$ and $i\nabla$. Denoting the corresponding linear
corrections by $\delta(\cdots)$, we have%
\begin{align}
&  \delta\left[  \frac{(\vec{p}\cdot\vec{G}_{e})(\hat{K}\cdot\vec{V}%
_{i0})-(\vec{p}\cdot\vec{G}_{i})(\hat{K}\cdot\vec{V}_{e0})}{\vec{p}\cdot
(\vec{G}_{e}-\vec{G}_{i})}\right] \nonumber\\
&  \approx(\vec{k}\cdot\vec{V}_{i0})\delta\left[  \frac{\vec{p}\cdot\vec
{G}_{e}}{\vec{p}\cdot(\vec{G}_{e}-\vec{G}_{i})}\right]  -(\vec{k}\cdot\vec
{V}_{e0})\delta\left[  \frac{\vec{p}\cdot\vec{G}_{i}}{\vec{p}\cdot(\vec{G}%
_{e}-\vec{G}_{i})}\right] \nonumber\\
&  +\frac{(\vec{k}\cdot\vec{G}_{e1})(\delta\hat{K}\cdot\vec{V}_{i0})-(\vec
{k}\cdot\vec{G}_{e1})(\delta\hat{K}\cdot\vec{V}_{i0})}{\vec{k}\cdot(\vec
{G}_{e1}-\vec{G}_{i1})}. \label{del_pervyj}%
\end{align}
According to Eq.~(\ref{Kq}), in the last term of Eq.~(\ref{del_pervyj}) we
have $\delta\hat{K}=-i\nabla$. Presuming a uniform magnetic field, $\vec
{B}=B\hat{b}$, expressing $\vec{V}_{e,i0}$ in terms of $\vec{U}_{0}$ according
to Eqs.~(\ref{V_0_II_via_U}), (\ref{V_0_perp_via_U}), and using
Eqs.~(\ref{proma}), (\ref{proma_2}), we obtain%
\begin{align}
&  \frac{(\vec{k}\cdot\vec{G}_{e1})(\delta\hat{K}\cdot\vec{V}_{i0})-(\vec
{k}\cdot\vec{G}_{e1})(\delta\hat{K}\cdot\vec{V}_{i0})}{\vec{k}\cdot(\vec
{G}_{e1}-\vec{G}_{i1})}\nonumber\\
&  =\frac{i}{\kappa_{e}+\kappa_{i}}\left[  \frac{(\kappa_{i}-\kappa_{e}%
)\nabla\cdot\vec{U}_{0}}{1+\psi_{k}}+\kappa_{e}\kappa_{i}\hat{b}\cdot
(\nabla\times\vec{U}_{0})\right]  . \label{term_prop_delta_K}%
\end{align}
Recall that $\nabla\cdot\vec{U}_{0}$ can be expressed in terms of $\nabla
n_{0}$ according to Eq.~(\ref{div_U_0_via_grad}). In this sense, the term
proportional to $\nabla\cdot\vec{U}_{0}$ can be considered as an additional
small contributor to the GD instability driving. The last vortex term
$\propto\nabla\times\vec{U}_{0}$ is unrelated to any density gradients, but it
may also contribute to $\gamma_{\vec{k}}$.

Next we turn to the combination of the two first terms in the RHS of
Eq.~(\ref{del_pervyj}). The difference between the two fractions $\left.
\vec{p}\cdot\vec{G}_{e}\right/  \vec{p}\cdot(\vec{G}_{e}-\vec{G}_{i})$ and
$\left.  \vec{p}\cdot\vec{G}_{i}\right/  \vec{p}\cdot(\vec{G}_{e}-\vec{G}%
_{i})$ equals a constant value of $1$, so that their linear
perturbations are equal and we have
\begin{subequations}
\begin{eqnarray}
&&(\vec{k}\cdot\vec{V}_{i0})\ \delta\left[  \frac{\vec{p}\cdot\vec{G}_{e}}{\vec
{p}\cdot(\vec{G}_{e}-\vec{G}_{i})}\right]\nonumber\\
&-&(\vec{k}\cdot\vec{V}_{e0})\ \delta\left[
\frac{\vec{p}\cdot\vec{G}_{i}}{\vec{p}\cdot(\vec{G}_{e}-\vec
{G}_{i})}\right]  =-\beta(\vec{k}\cdot\vec{U}_{0}), \label{via_beta}%
\end{eqnarray}
\end{subequations}
where
\begin{subequations}
\label{betas}%
\begin{align}
\beta &  \equiv\delta\left[  \frac{\vec{p}\cdot\vec{G}_{e}}{\vec{p}\cdot
(\vec{G}_{e}-\vec{G}_{i})}\right]  =\delta\left[  \frac{\vec{p}\cdot\vec
{G}_{i}}{\vec{p}\cdot(\vec{G}_{e}-\vec{G}_{i})}\right]  =\beta_{\mathrm{FB}%
}+\beta_{\mathrm{GD}},\label{beta_full}\\
\beta_{\mathrm{FB}}  &  \approx\frac{(\vec{k}\cdot\vec{G}_{e1})(\vec{k}%
\cdot\delta\vec{G}_{i})-(\vec{k}\cdot\vec{G}_{i1})(\vec{k}\cdot\delta\vec
{G}_{e})}{[\vec{k}\cdot(\vec{G}_{e1}-\vec{G}_{i1})]^{2}},\label{beta_FB}\\
\beta_{\mathrm{GD}}  &  \approx\frac{(\vec{k}\cdot\vec{G}_{e1})(\vec{G}%
_{i1}\cdot\delta\vec{p})-(\vec{k}\cdot\vec{G}_{i1})(\vec{G}_{e1}\cdot
\delta\vec{p})}{[\vec{k}\cdot(\vec{G}_{e1}-\vec{G}_{i1})]^{2}}.
\label{beta_GD}%
\end{align}
\end{subequations}
Here the coefficient $\beta_{\mathrm{FB}}$ includes particle-inertia
contributions responsible for the FB instability, while $\beta_{\mathrm{GD}}$
includes the background density gradients responsible for the GD instability.

First, we calculate the FB-instability driving term, $-\beta_{\mathrm{FB}%
}(\vec{k}\cdot\vec{U}_{0})$. According to Eqs.~(\ref{G}) and
Eq.~(\ref{tilde_kappa}), we obtain
\begin{align*}
\vec{k}\cdot\delta\vec{G}_{e}  &  =-\ \frac{\kappa_{e}k_{\perp}^{2}}{B}\left[
\frac{1-\kappa_{e}^{2}}{(1+\kappa_{e}^{2})^{2}}+\frac{k_{\parallel}^{2}%
}{k_{\perp}^{2}}\right]  \frac{\Omega_{\vec{k}1}^{(e)}}{\nu_{e}},\\
\vec{k}\cdot\delta\vec{G}_{i}  &  =\frac{\kappa_{i}k_{\perp}^{2}}{B}\left[
\frac{1-\kappa_{i}^{2}}{(1+\kappa_{i}^{2})^{2}}+\frac{k_{\parallel}^{2}%
}{k_{\perp}^{2}}\right]  \frac{\Omega_{\vec{k}1}^{(i)}}{\nu_{i}}.
\end{align*}
Using Eq.~(\ref{Omega_k_1}), we have $\vec{k}\cdot\vec{G}_{e,i1}$,%
\begin{align}
\vec{k}\cdot\vec{G}_{e1} & =-\ \frac{\vec{k}\cdot\left(  \vec{G}_{e1}-\vec
{G}_{i1}\right)  \Omega_{\vec{k}1}^{(e)}}{\vec{k}\cdot\vec{U}_{0}},\nonumber\\
\vec{k}\cdot\vec{G}_{i1} & =-\ \frac{\vec{k}\cdot\left(  \vec{G}_{e1}-\vec
{G}_{i1}\right)  \Omega_{\vec{k}1}^{(i)}}{\vec{k}\cdot\vec{U}_{0}},\nonumber
\end{align}
where according to Eq.~(\ref{proma_2}),%
\[
\vec{k}\cdot(\vec{G}_{e1}-\vec{G}_{i1})=\frac{i\kappa_{i}\kappa_{e}\left(
\kappa_{e}+\kappa_{i}\right)  \left(  1+\psi_{\vec{k}}\right)  k_{\perp}^{2}%
}{\left(  1+\kappa_{e}^{2}\right)  \left(  1+\kappa_{i}^{2}\right)  B}.
\]
As a result, we obtain from Eq.~(\ref{beta_FB})%
\begin{align}
&  -\beta_{\mathrm{FB}}(\vec{k}\cdot\vec{U}_{0})\nonumber\\
&  =\frac{i\Omega_{\vec{k}1}^{(e)}\Omega_{\vec{k}1}^{(i)}}{\left(  \kappa
_{e}+\kappa_{i}\right)  \left(  1+\psi_{\vec{k}}\right)  }%
\label{beta_FB_final}\\
&  \times\!\!\left[  \frac{(1-\kappa_{i}^{2})(1+\kappa
_{e}^{2})  }{(1+\kappa_{i}^{2})  \kappa_{e}\nu_{i}}
+\frac{(1-\kappa_{e}^{2})(1+\kappa_{i}^{2})
}{(1+\kappa_{e}^{2})  \kappa_{i}\nu_{e}}+\left(  \frac{\kappa
_{e}}{\nu_{e}}+\frac{\kappa_{i}}{\nu_{i}}\right)  \frac{k_{\parallel}^{2}%
}{k_{\perp}^{2}}\right]\!\!  .\nonumber
\end{align}
Here the first and second terms in the square bracket originate
from the ion and electron inertia, respectively. These terms
correspond to particle oscillation in the perpendicular to
$\vec{B}_{0}$ plane. Notice that for $\kappa_{e}>1$, the
electron inertia terms becomes negative, meaning that it
opposes the FB instability. The physical nature of this
electron-inertia stabilization is fully analogous to that for
the ion inertia above the magnetization boundary,
$\kappa_{i}>1$ \citep{DimOppen2004:ionthermal1}. The last term
in the square bracket of Eq.~(\ref{beta_FB_final}) includes the
effects of the electron ($\kappa_{e}/\nu_{e}$) and ion
($\kappa_{i}/\nu_{i}$) inertia in the parallel to $\vec{B}_{0}$
direction. The second and third terms in the square bracket are
usually neglected, but the conditions for such neglect have not
being properly analyzed and justified. We do this below.

Similarly, we calculate the term $-\beta_{\mathrm{GD}}(\vec{k}\cdot\vec{U}%
_{0})$ that describes the local gradient drift instability or
stabilization. According to Eq.~(\ref{p}),
$\delta\vec{p}=-i\nabla n_{0}/n_{0}$, and we obtain from
Eq.~(\ref{beta_GD}),
\begin{align}
&  -\beta_{\mathrm{GD}}(\vec{k}\cdot\vec{U}_{0})=\frac{i(1+\kappa_{e}%
^{2})(1+\kappa_{i}^{2})(\vec{k}\cdot\vec{U}_{0})}{\kappa_{e}\kappa_{i}%
(\kappa_{e}+\kappa_{i})(1+\psi_{k})^{2}k_{\perp}^{2}}\nonumber\\
&  \times\left\{  (\kappa_{e}-\kappa_{i})\left(  \vec{k}_{\parallel}\cdot
\frac{\nabla_{\parallel}n_{0}}{n_{0}}-\vec{k}_{\perp}\cdot\frac{\nabla_{\perp
}n_{0}}{n_{0}}\ \frac{k_{\parallel}^{2}}{k_{\perp}^{2}}\right)  \right.
\label{beta_GD_final}\\
&  -\left.  \left[  1+(1+\kappa_{i}\kappa_{e})\frac{k_{\parallel}^{2}%
}{k_{\perp}^{2}}\right]  (\vec{k}\times\hat{b})\cdot\frac{\nabla_{\perp}n_{0}%
}{n_{0}}\right\}  .\nonumber
\end{align}

To summarize, we obtain for the combined FB and GD linear growth rate the
general expression%
\begin{equation}
\gamma_{\vec{k}}=\gamma_{\mathrm{FB}}+\gamma_{\mathrm{GD}}-\gamma_{T}-2\alpha
n_{0}, \label{gamma_total}%
\end{equation}
where%
\begin{align}
\gamma_{\mathrm{FB}}  &  =-\ \frac{\Omega_{\vec{k}1}^{(e)}\Omega_{\vec{k}%
1}^{(i)}}{\kappa_{e}\left(  1+\psi_{\vec{k}}\right)  }\left[  \frac{\left(
1-\kappa_{i}^{2}\right)  \left(  1+\kappa_{e}^{2}\right)  }{\left(
1+\kappa_{i}^{2}\right)  \kappa_{e}\nu_{i}}\right. \nonumber\\
&  \left.  +\frac{\left(  1-\kappa_{e}^{2}\right)  \left(  1+\kappa_{i}%
^{2}\right)  }{\left(  1+\kappa_{e}^{2}\right)  \kappa_{i}\nu_{e}}+\left(
\frac{\kappa_{e}}{\nu_{e}}+\frac{\kappa_{i}}{\nu_{i}}\right)  \frac
{k_{\parallel}^{2}}{k_{\perp}^{2}}\right]  , \label{gamma_FB}%
\end{align}%
\begin{align}
&  \gamma_{\mathrm{GD}}=\frac{(1+\kappa_{e}^{2})(1+\kappa_{i}^{2})(\vec
{k}\cdot\vec{U}_{0})}{\kappa_{e}\kappa_{i}(\kappa_{e}+\kappa_{i})(1+\psi
_{k})^{2}k_{\perp}^{2}}\nonumber\\
&  \times\left\{  (\kappa_{e}-\kappa_{i})\left(  \vec{k}_{\parallel}\cdot
\frac{\nabla_{\parallel}n_{0}}{n_{0}}-\vec{k}_{\perp}\cdot\frac{\nabla_{\perp
}n_{0}}{n_{0}}\ \frac{k_{\parallel}^{2}}{k_{\perp}^{2}}\right)  \right.
\nonumber\\
&  -\left.  \left[  1+(1+\kappa_{i}\kappa_{e})\frac{k_{\parallel}^{2}%
}{k_{\perp}^{2}}\right]  (\vec{k}\times\hat{b})\cdot\frac{\nabla_{\perp}n_{0}%
}{n_{0}}\right\}  , \label{Gamma_GD}%
\end{align}%
\begin{equation}
\gamma_{T}=\frac{k_{\perp}^{2}\left[  1+(1+\kappa_{e}^{2})k_{\parallel}%
^{2}/k_{\perp}^{2}\right]  \left[  1+(1+\kappa_{i}^{2})k_{\parallel}%
^{2}/k_{\perp}^{2}\right]  (T_{e}+T_{i})}{(\kappa_{e}+\kappa_{i})(1+\psi
_{\vec{k}})eB}, \label{gamma_T}%
\end{equation}
and the first-order shifted frequencies $\Omega_{\vec{k}1}^{(e,i)}$ are given
by Eqs.~(36) and (39) from \citet{Dimant:Magnetosphere2011_Budget}.

For magnetized electrons at a sufficiently high altitude where
$\kappa_{e}\gg1$ and
$\psi_{\perp}\equiv(\kappa_{e}\kappa_{i})^{-1}\ll
\nu_{e}/\nu_{i}\simeq10$ hold together (see below), presuming
$|k_{\parallel}|/k_{\perp
}\lesssim\kappa_{e}^{-1}\lll\kappa_{i}^{-1}$, we obtain
\begin{equation}
\Omega_{\vec{k}1}^{(i)}\approx\frac{\vec{k}\cdot\vec{U}_{0}}{1+\psi_{\vec{k}}%
},\qquad\Omega_{\vec{k}1}^{(e)}\approx-\psi_{\vec{k}}(1+\kappa_{i}^{2}%
)\Omega_{\vec{k}1}^{(i)} \label{Omegi_reduced}%
\end{equation}
so that%
\begin{equation}
\gamma_{\mathrm{FB}}\approx\frac{\psi_{\vec{k}}\left(  1-\kappa_{i}%
^{2}\right)  (\Omega_{\vec{k}1}^{(i)})^{2}}{\left(  1+\psi_{\vec{k}}\right)
\nu_{i}},\qquad\gamma_{T}\approx\frac{\psi_{\vec{k}}k_{\perp}^{2}C_{s}^{2}%
}{(1+\psi_{\vec{k}})\nu_{i}}, \label{gamma_FB_reduced}%
\end{equation}%
\begin{equation}
\gamma_{\mathrm{GD}}=\frac{(1+\kappa_{i}^{2})\Omega_{\vec{k}1}^{(i)}}%
{\kappa_{i}(1+\psi_{\vec{k}})k_{\perp}^{2}}\left[  \kappa_{e}\vec
{k}_{\parallel}\cdot\frac{\nabla_{\parallel}n_{0}}{n_{0}}-(\vec{k}\times
\hat{b})\cdot\frac{\nabla_{\perp}n_{0}}{n_{0}}\right]  ,
\label{gamma_GD_reduced}%
\end{equation}
where $C_{s}^{2}\equiv(T_{e}+T_{i})/m_{i}$.
Equation~(\ref{gamma_FB_reduced}) shows that for fluid
particles ion inertia is a destabilizing factor
($\gamma_{\mathrm{FB}}>0$) for all altitudes below the
magnetization boundary defined by $\kappa_{i}=1$. For higher
altitudes with $\kappa_{i}>1$, ion inertia is a stabilizing
factor, $\gamma_{\mathrm{FB}}\leq0$. The physical nature of
this peculiar feature has been discussed in
\citet{DimOppen2004:ionthermal1}. Note that the multipliers
$\left(  1+\psi_{\vec{k}}\right)  $ in
Eqs.~(\ref{Omegi_reduced}) to (\ref{gamma_GD_reduced}) differ
from the corresponding multipliers in the well-known
expressions \citep{Fejer:Theory84} with the traditional
definition of $\psi$ by a term
$\psi_{\perp}\kappa_{i}^{2}=\kappa_{i}/\kappa_{e}\equiv\theta
_{0}^{2}\simeq1.8\times10^{-4}$, that can be neglected.

Now we discuss some of the conventionally neglected terms in
the linear dispersion relation. The second and third terms in
the square bracket of Eq.~(\ref{beta_FB_final}) have never been
taken into account. In the case under consideration, the ratio
of the second term to the first one is
$\nu_{i}/(\kappa_{i}\kappa_{e}\nu_{e})=\psi_{\perp}(\nu_{i}/\nu_{e})$.
In the lower ionosphere, we have $\nu_{i}/\nu_{e}\simeq0.1$, so
that the electron inertia in the perpendicular to $\vec{B}_{0}$
plane can be neglected provided $\psi_{\perp}\ll10$. This
condition is fulfilled well above the 90~km altitude, i.e., in
essentially within the entire altitude range of a possible FB
instability development. At lower altitudes (the upper
\textit{D} region), the electron inertia can become an
additional stabilizing factor for the electron thermal driven
instability \citep{Dimant:Physical97}.

For the third term in the square brackets of Eq.~(\ref{beta_FB_final}), the
ratio of its two sub-terms is $(\kappa_{e}/\nu_{e})/(\kappa_{i}/\nu
_{i})=(m_{i}/m_{e})(\nu_{i}/\nu_{e})^{2}\simeq550$, so that in the parallel to
$\vec{B}_{0}$ direction it is the electron inertia term $\propto\kappa_{e}/\nu
_{e}$ that largely dominates. Then the ratio of the third term to the first
one becomes $(\nu_{e}/\nu_{i})k_{\parallel}^{2}/k_{\perp}^{2}$ $\simeq
10(k_{\parallel}/k_{\perp})$. It is of order unity or larger for the wave
aspect angles, $\theta\equiv\arctan(k_{\parallel}/k_{\perp})\gtrsim20^{\circ}%
$. Thus, particle inertia in the parallel to $\vec{B}_{0}$
direction, which is largely due to electrons, can only affect
waves whose wavevectors lie well off the perpendicular plane,
usually outside the angular domain of linear instabilities.
Such waves are heavily damped and hardly play a role in any
linear or non-linear processes. This means that for waves in
the lower ionosphere the last term in the square bracket of
Eq.~(\ref{beta_FB_final}) is probably never of importance.

\begin{acknowledgments}
This work was supported by National Science Foundation
Ionospheric Physics Grants No.~ATM-0442075, ATM-0819914 and
ATM-1007789. The authors thank Dr. V.~G.~Merkin for useful
discussions of conductance issues in global MHD models.
\end{acknowledgments}


\end{article}
\end{document}